\documentclass{iopart}
\usepackage[dvips]{graphicx}
\usepackage{iopams}
\def \Heff{\mathcal{H}_{\mathrm{eff}}}
\def \aver#1{\left\langle#1\right\rangle}
\def \averc#1{\left\langle#1\right\rangle_{\mathrm{conn}}}

\def \published#1 {\vspace{28pt plus 10pt minus 18pt}
     \noindent{\small\tt Published in: {\rm #1}\par}}

\begin{document}

\review[Scattering, reflection and impedance of chaotic waves]{Scattering,
reflection and impedance of waves in chaotic and disordered systems with
absorption}
\author{Y V Fyodorov$^1$, D V Savin$^2$ and H-J Sommers$^2$ }
\address{$^1$\,School of Mathematical Sciences, University of
Nottingham, Nottingham NG7 2RD, UK}
\address{$^2$\,Fachbereich Physik, Universit\"at Duisburg-Essen, 45117 Essen, Germany}

\begin{abstract} \\
We review recent progress in analysing wave scattering in systems with both
intrinsic chaos and/or disorder and internal losses, when the scattering
matrix is no longer unitary. By mapping the problem onto a nonlinear
supersymmetric $\sigma$--model, we are able to derive closed form analytic
expressions for the distribution of reflection probability in a generic
disordered system. One of the most important properties resulting from such
an analysis is statistical independence between the phase and the modulus of
the reflection amplitude in every perfectly open channel. The developed
theory has far-reaching consequences for many quantities of interest,
including local Green functions and time delays. In particular, we point out
the role played by absorption as a sensitive indicator of mechanisms behind
the Anderson localisation transition. We also provide a random-matrix based
analysis of $S$-matrix and impedance correlations for various symmetry
classes as well as the distribution of transmitted power for systems with
broken time-reversal invariance, completing previous works on the subject.
The results can be applied, in particular, to the experimentally accessible
impedance and reflection in a microwave or an ultrasonic cavity attached to
a system of antennas.
\end{abstract}


\published{J. Phys. A: Math. Gen. \textbf{38} (2005) 10731--10760}

\section{Introduction}

Propagation of electromagnetic or ultrasonic waves in billiards
\cite{Stoeckmann}, compound-nucleus reactions \cite{Verbaarschot1985},
scattering of light in random media and transport of electrons through
quantum dots \cite{Beenakker1997,Alhassid2000} share at least one feature in
common: in all these situations one deals with an open wave-chaotic system
studied by means of a scattering experiment, see figure~1 for an
illustration. Here, we have a typical transport problem where the
fundamental object of interest is the scattering matrix $S$, which relates
linearly the amplitudes of incoming and outgoing fluxes. However, under real
laboratory conditions there is a number of different sources which cause
that a part of the flux gets irreversibly lost or dissolved in the
environment. As a result, we encounter absorption and have to handle the
$S$-matrix, which is no longer unitary. Statistics of different scattering
observables in the presence of absorption are nowadays under intensive
experimental and theoretical investigations, starting from early experiments
on reflection and energy correlations of the $S$-matrix
\cite{Doron1990,Lewenkopf1992i}. More recently, total cross-sections
\cite{Schaefer2003}, distributions of reflection \cite{Mendez-Sanchez2003}
and transmission \cite{Schanze2005} coefficients as well as that of the
complete $S$ matrix \cite{Kuhl2005} in microwave cavities, properties of
resonance widths \cite{Barthelemy2005b} in such systems at room
temperatures, dissipation of ultrasonic energy in elastodynamic billiards
\cite{Lobkis2003}, and fluctuations in microwave networks \cite{Hul2004}
became experimentally available. Theoretically, statistics of reflection,
delay times and related quantities were considered first in the strong
\cite{Kogan2000} and then weak \cite{Beenakker2001} absorption limits at
perfect coupling and very recently at arbitrary absorption and coupling for
several symmetry classes
\cite{Savin2003i,Fyodorov2003i,Savin2004,Rozhkov2003,Rozhkov2004,
Fyodorov2004ii,Savin2005,Martinez-Mares2005,Fyodorov2004iii}.

\begin{figure}
\mbox{\ }\hfill\includegraphics[width=0.8\textwidth]{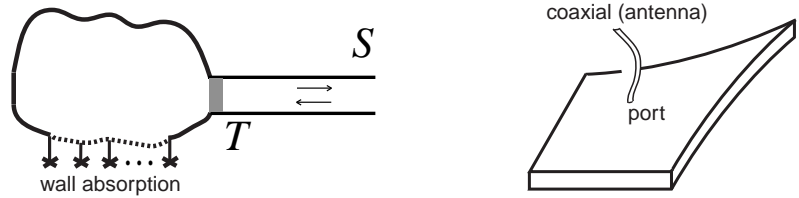}\hfill
\caption{A sketch of a typical experimental setup with microwave billiards.
A flat chaotic cavity is feeded with microwaves through an attached coaxial
cable (i.e. a scattering channel). On average, $1-T$ part of the incoming
flux, where $T\leq1$ is the so-called transmission coefficient, is reflected
back directly from the cable-cavity interface (port) without exciting
long-lived resonances in the cavity. If the cavity is thin enough then only
a transverse electric wave can propagate inside. The electric field has only
a vertical component, which is uniform in vertical direction and distributed
nontrivially in the plane. Therefore, there is a voltage between plates as
well as a current due to the in-plane magnetic field. The impedance is a
quantity which relates linearly the port voltage to the port current.
Fluctuations of eigenmodes and eigenfrequencies result in fluctuations of
the impedance or $S$-matrix, as the driving frequency or port position is
changed.}
\end{figure}

Another insight to the same problem comes from looking at it not from the
``outside'', but rather from the ``inside''. Then the prime object of
interest turns out to be the impedance $Z$ relating linearly voltages to
currents at the system input \cite{Hemmady2005,Hemmady2005i}, see figure~1.
By proper taking into account the wave nature of the current
\cite{Zheng2004a,Zheng2004b} the cavity impedance can be seen as an
electromagnetic analogue of Wigner's reaction matrix of the scattering
theory. This can be easily understood qualitatively through the well-known
equivalence of the two-dimensional Maxwell equations to the Schr\"odinger
equation, the role of the wave function being played by the electromagnetic
field (the voltage in our case). Then the definition of the impedance
becomes formally similar to the definition of the reaction matrix (which
relates linearly the scattering wave function to its normal derivative on
the boundary). The impedance is, therefore, related to the local Green's
function of the closed cavity and fluctuates strongly due to chaotic
internal dynamics.

The imaginary part of the local Green's function (which is
proportional to the real part of $Z$) is well known as the local
density of states (LDoS) and has a long story of studies in
disordered electronic systems, see \cite{Mirlin2000} for a recent
review. Actually, a closely related quantity emerges in the
context of spectra of complex atoms and molecules where it has the
meaning of the total cross-section of indirect photoabsorption
\cite{Fyodorov1998i} (see also \cite{Sokolov1997ii,Gorin2004}). It
also appears in studies on spontaneous light emission by atoms
placed in chaotic cavities \cite{Misirpashaev1997}. As to the real
part of the Green's function, it seems to have no direct physical
meaning in mesoscopics while it has the meaning of reactance in
electromagnetics, where both real and imaginary parts are
experimentally accessible.

In this review we discuss an approach developed recently by us in
short communications  \cite{Fyodorov2004ii,Savin2005,Savin2005i}
which treats both inside and outside aspects of the problem on
equal footing. In this capacity it provides a uniform and deeper
understanding of various results on absorptive scattering obtained
earlier in \cite{Savin2003i,Fyodorov2003i,Savin2004}.
 In particular, the method allows one to study very efficiently the
distribution of the local Green's function (complex impedance) at arbitrary
absorption and to relate it to that of reflection, thus linking somewhat
complementary experiments \cite{Kuhl2005} and \cite{Hemmady2005} together.
Although calculations are most explicit and simple for fully chaotic
systems, when one can rely upon the random matrix theory (RMT), our method
actually has relevance in a much broader context beyond RMT that involves
many interesting aspects of disordered mesoscopic systems with absorption,
including effects of the Anderson localisation.  From that point of view,
the method opens an attractive possibility to look at some long-standing
problems (e.g. statistics of time delays) from a different perspective, see
\cite{Ossipov2005} as well as subsections 4.1.6 and 4.1.7 below.

\section{Reflection, time delays and resonance spectrum}

In this section, we provide a short description of the scattering
approach to the problem. The resonance energy dependence of
observables becomes explicit in the well-known Hamiltonian
approach to quantum scattering, which was developed first in the
context of  nuclear physics
\cite{Verbaarschot1985,Livsic1957,Mahaux,Sokolov1989} and can be
easily adopted for models emerging in quantum chaotic scattering
and mesoscopic physics, see e.g.
\cite{Alhassid2000,Lewenkopf1991,Fyodorov1997,Dittes2000} for
reviews. This framework turns out to be also most suited to take a
finite absorption into account.  The starting point is the
following fundamental relation between the resonance part of the
scattering matrix and the Wigner's reaction matrix $K$:
\begin{equation}\label{S}
S(E) = \frac{1-\rmi K(E)}{1+\rmi K(E)}\,, \qquad
K(E)=\case12V^{\dag}(E-H)^{-1}V\,.
\end{equation}
The Hermitian Hamiltonian $H$ of the \emph{closed} system gives rise to $N$
real energy levels (eigenfrequencies). Those are coupled to $M$ continuum
channels via the $N{\times}M$ matrix $V$ of coupling amplitudes $V^c_n$
($n=1\ldots N$, $c=1\ldots M$), and as a result are converted to $N$ complex
{\it resonances}. To see this, we expand (\ref{S}) in a Taylor series in $K$
and, after regrouping the terms, bring the resulting expression to another
well known form
\begin{equation}\label{Heff}
S(E) = 1-\rmi V^{\dag}\frac{1}{E-\Heff}V\,, \qquad
\Heff=H-\case\rmi2VV^{\dag}\
\end{equation}
for the $S$-matrix. The effective Hamiltonian $\Heff$ emerging here
characterises the \emph{open} system and is the non-Hermitian counterpart of
$H$. The factorized structure of the anti-Hermitian part is necessary to
ensure the unitarity of $S$. The coupling amplitudes $V$ change very slowly
with the energy (far from the channels thresholds) and one can safely
consider them to be energy-independent. In such a resonance approximation
the complex eigenvalues $\mathcal{E}_n=E_n-\case\rmi2\Gamma_n$ of $\Heff$,
with the energies $E_n$ and escape widths $\Gamma_n>0$, are the only
singularities of the $S$ matrix in the complex energy plane. As required by
causality \cite{Nussenzveig}, they are located in the lower half plane and
correspond to the long-lived resonance states formed on the intermediate
stage of a scattering process.
 The corresponding (left and right) eigenvectors form the
so-called bi-orthogonal system. Recent discussion concerning
applicability of the effective Hamiltonian approach to potential
scattering problems (like those with cavities) can be found in
\cite{Pichugin2001,Stoeckmann2002i,Savin2003,Barthelemy2005a}.

Flux conservation requires $S(E)$ to be unitary at the real values of $E$.
It is useful to define at real $\Omega$ the following matrix
\begin{eqnarray}\label{Romega}
\fl R_{\Omega}(E) \equiv S^{\dag}(E-\case12\Omega)S(E+\case12\Omega)
=1+\rmi\Omega V^{\dag} \frac{1}{E-\case12\Omega-\Heff^{\dag}}
\frac{1}{E+\case12\Omega-\Heff}V\,.
\end{eqnarray}
The second equality here results from the substitution of (\ref{Heff}) and
then making use of the identity $\rmi VV^{\dag}+\Omega =
(E+\case12\Omega-\Heff)-(E-\case12\Omega-\Heff^{\dag})$ which leads to
cancellations of the cross-terms. Expression (\ref{Romega}) tends to unity
as $\Omega\to0$ and it is the Wigner-Smith time-delay matrix
\cite{Wigner1955,Smith1960} $Q(E)=-\rmi S^{\dag}(E)\case{\rmd}{\rmd E}
S(E)=-\rmi\case{\rmd}{\rmd\Omega}R_{\Omega}(E)|_{\Omega=0}$ (we put
$\hbar=1$) which determines the unitarity deficit of $R_{\Omega}$ to the
linear order \cite{Lehmann1995b}:
\begin{eqnarray}\label{Q}
\fl R_{\Omega}(E) \simeq 1+\rmi\Omega Q(E) + O(\Omega^2)\,, \qquad
Q(E)=V^{\dag}\frac{1}{(E-\Heff)^{\dag}}\frac{1}{E-\Heff}V\,.
\end{eqnarray}
Such a factorized representation \cite{Sokolov1997} for the time-delay
matrix (which contains no longer the energy derivative) is a consequence of
the resonance approximation considered. It serves to make a connection of
time delays to the resonance spectrum most explicit. The matrix element
$Q^{cc'}=(b^{\dag}b)^{cc'}$ may be physically interpreted as the scalar
product (or ``overlap'') between the internal parts $b^c=(E-\Heff)^{-1}V^c$
of the scattering wave functions for waves incident in the channels $c$ and
$c'$, respectively \cite{Sokolov1997}. In particular, the mean time delay in
the channel $c$ given by the diagonal element $Q^{cc}$ coincides in such an
approximation with the dwell time given by the norm of $b^c$. One should
distinguish generally between $Q^{cc}$ and the so-called \emph{proper}
time-delays $q_c$ (eigenvalues of $Q$). Taking their sum, one comes to a
weighted mean time delay characteristic $Q_w=\frac{1}{M}\tr
Q=\frac{1}{M}\sum_c Q^{cc} =\frac{1}{M}\sum_c q_c$, the so-called Wigner's
time delay which is known, see e.g.
\cite{Lyuboshitz1977,Lehmann1995b,Fyodorov1997}, to be determined by the
energy derivative of the total scattering phaseshift:
$Q_w=-\case{\rmi}{M}\frac{\partial}{\partial E}\log\det{S}$. Diverse aspects
of delay times in quantum chaotic scattering \cite{Fyodorov1997} as well as
in a general quantum mechanical context \cite{Muga2000,deCarvalho2002} can
be found in the cited literature and references therein.

As is well known, statistics of spectra of {\it closed} quantum systems with
chaotic classical counterparts are to a large extent universal and
independent of their microscopic nature. This remarkable universality
provides one with a possibility to use the RMT \cite{Mehta2} for an adequate
description of many physical properties of such systems \cite{Guhr1998}.
According to the general paradigm we replace the actual Hermitian part $H$
of the effective Hamiltonian (\ref{Heff}) with a random Hermitian matrix $H$
taken from one of the three canonical Wigner-Dyson's ensembles labelled by
the symmetry index $\beta$ according to the symmetry of the system under
consideration. The Gaussian orthogonal (GOE, $\beta=1$ and $H$ real
symmetric) and unitary (GUE, $\beta=2$ and $H$ Hermitian) ensembles stand
for systems with preserved or fully broken time-reversal symmetry (TRS),
respectively. The remaining Gaussian symplectic ensemble (GSE, $\beta=4$ and
$H$ self-dual quaternion) is relevant for description of time-reversal
systems with strong spin-orbit scattering. The limit of large $N\to\infty$
is supposed to be finally taken. Then eigenvalues are distributed on the
finite interval according to Wigner's semicircle law, which determines
locally the mean level spacing $\Delta$. The most appealing feature of the
RMT approach is that quantities related to spectral fluctuations when
expressed in units of $\Delta$ (``unfolding'') do not depend on microscopic
details (i.e. the particular form of the distribution of $H$ or the profile
of $\Delta$) and become uniform throughout the whole spectrum
\cite{Guhr1998}. For practical reasons we thus restrict ourselves to
considering fluctuations at the center of the spectrum ($E=0$) only.
Similarly, the results turn out to be also independent of particular
statistical assumptions on coupling amplitudes $V^{c}_{n}$ as long as $M\ll
N$ \cite{Lehmann1995a,Lehmann1995b}. The amplitudes may be chosen as fixed
\cite{Verbaarschot1985} or random \cite{Sokolov1989} variables and enter
final expressions only in combinations known as transmission coefficients
(also sometimes called sticking probabilities)
\begin{equation}\label{T}
T_c\equiv1-|\overline{S^{cc}}|^2=\frac{4\kappa_c}{(1+\kappa_c)^2}\,, \qquad
\kappa_c=\frac{\pi\|V^c\|^2}{2N\Delta}
\end{equation}
where $\overline{S^{cc}}$ stands for the average (optical) $S$ matrix. The
transmission coefficients are assumed to be input parameters of the theory,
the cases $T_c\ll1$ and $T_c=1$ corresponding to an almost closed or
perfectly open channel ``c'', respectively.

Absorption is usually seen as a dissipation process, which evolves
exponentially in time. Strictly speaking, different spectral components of
the field may have different dissipation rates. However, frequently this
rather weak energy dependence can easily be neglected as long as
\emph{local} fluctuations on much finer energy scale $\sim\Delta$ are
considered. As a result, all the resonances acquire one and the same
absorption width $\Gamma>0$ additionally to their escape widths $\Gamma_n$.
The dimensionless phenomenological parameter $\gamma\equiv2\pi\Gamma/\Delta$
characterizes then the absorption strength, with $\gamma\ll1$ or
$\gamma\gg1$ corresponding to the weak or strong absorption limit,
respectively. Microscopically, it can be modelled by means of a huge number
of weakly open parasitic channels \cite{Lewenkopf1992i,Brouwer1997ii} or by
coupling to a very complicated background with almost continuous spectrum
\cite{Savin2003i}, see also \cite{Sokolov2005}. In microwave billiards such
an approximation is frequently very good to account for uniform Ohmic losses
which happen everywhere in non-perfectly conducting walls. However, in some
experimentally relevant situations as, e.g., complex reverberant structures
\cite{Lobkis2000} or even microwave cavities at room temperature
\cite{Barthelemy2005b,Barthelemy2005a} an approximation of uniform
absorption may break down, and one should take into account instead
localized-in-space losses which will result in different broadenings of
different modes. The latter are easily incorporated in the model by treating
them as if induced by additional scattering channels, see e.g.
\cite{Rozhkov2003}. An alternative scheme of treating localised-in-space
surface absorption is discussed in \cite{Martinez-Mares2005}. (Discussion of
a formal theory of scattering for complex absorbing potentials can be found
in \cite{Muga2004}.)

Operationally, the \emph{uniform} absorption can equivalently be
taken into account by a purely imaginary shift of the scattering
energy $E\to E+\case\rmi2\Gamma\equiv E_{\gamma}$, so that the
$S$-matrix $S_{\gamma}(E)\equiv S(E_{\gamma})$ becomes subunitary.
The reflection matrix $R_{\gamma}=S_{\gamma}^{\dag}S_{\gamma}$
provides then a natural measure of the mismatch between incoming
and outgoing fluxes. It can be obtained from $R_{\Omega}$,
(\ref{Romega}), by analytic continuation in $\Omega$ from a real
to the purely imaginary value $\Omega=\rmi\Gamma$, yielding
\cite{Savin2003i}:
\begin{eqnarray}\label{R} \fl R_{\gamma}(E) =
R_{\Omega=\rmi\Gamma}(E) = 1-\Gamma Q_{\gamma}(E) \,, \qquad
Q_{\gamma}(E)\equiv
V^{\dag}\frac{1}{(E_{\gamma}-\Heff)^{\dag}}\frac{1}{E_{\gamma}-\Heff}V\,.
\end{eqnarray}
This representation is valid at arbitrary value of $\Gamma$. In
the limit of small $\Gamma$ one can neglect the difference between
$Q_{\gamma}$ and $Q$, resulting in the approximate expression
\cite{Ramakrishna2000,Beenakker2001} $R_{\gamma}\simeq1-\Gamma Q$
following from (\ref{Q}). It is therefore tempting to keep for
$Q_{\gamma}$ the meaning of the time-delay matrix at finite
absorption as well (see, however, discussion in
\cite{Savin2003i}). By construction, $R_{\gamma}$ is a Hermitian
matrix, its positive \emph{reflection eigenvalues} $r_{c}=1-\Gamma
q_c\le1$ are related to proper time-delays $q_c$ (at finite
absorption).

Considering quantum scattering with no internal dissipation,
$\Gamma=0$, the $S$-matrix unitarity ensures that the reflection
coefficient $R_c=|S^{cc}|^2$ in any given channel is simply
related to the quantum mechanical probability to exit via any of
the remaining channels, known as the transmission probability:
$R_c=1-\sum_{b\ne c}|S^{bc}|^2$. For an absorptive system the last
equality is violated, but still the quantity $\tau_c=1-R_c$ can be
interpreted as the quantum mechanical probability that a particle
entering via a given channel never exits through the same channel.
Hence it was suggested to call $\tau_c$ the \emph{probability of
no return} (PNR) \cite{Fyodorov2003i}. In the particular case of a
single open channel fluctuations of the PNR $\tau$ arise solely
due to absorption (neglecting dissipation trivially results in
$\tau\equiv 0$). At weak absorption $\tau \approx \Gamma q$ , so
that the PNR is just simply proportional to the time-delay.

At last but not least, the matrix $Z\equiv\rmi K(E_{\gamma})$ has the
meaning of the normalized cavity impedance in such a setting, see
\cite{Zheng2004a,Zheng2004b} for further details.

\section{Correlation functions}

Any observable in chaotic resonance scattering exhibits strong fluctuations
over a smooth background as the scattering energy or other external
parameters are varied. Usually these variations occur on two essentially
different energy scales. This fact is conventionally taken into account by
decomposing fluctuating quantities into a mean part and a fluctuating part,
the former being understood as the result of spectral or (assumed to be
equivalent) ensemble average $\langle\cdots\rangle$. In this section we
consider statistics as determined by a two-point correlation function of the
fluctuating parts (frequently called a ``connected'' correlation function):
$\averc{AB}=\aver{AB}-\aver{A}\aver{B}$. We restrict ourselves below to the
cases of preserved ($\beta=1$) and broken ($\beta=2$) TRS (the symplectic
case $\beta=4$ proceeds along the same lines).

\subsection{Impedance}%
Let us start with considering the simplest case of the impedance
correlations. The problem can be fully reduced to that of spectral
correlations determined by the two-point cluster function
$Y_{2,\beta}(\omega)=\delta(\omega)-\Delta^2\aver{\rho(E_1)\rho(E_2)}_{\mathrm{conn}}$,
where $\omega=(E_2-E_1)/\Delta$ and $\rho(E)$ being the level density. It is
easy to satisfy oneself that for the mean impedance at $E=0$ holds
$\aver{Z^{ab}}=\kappa_{a}\delta^{ab}$. To calculate the energy correlation
function $C^{abcd}_Z(\omega)\equiv\averc{Z^{ab*}(E_1) Z^{cd}(E_2)}$, it is
instructive to write $Z^{ab}(E)$ in the eigenbasis of the closed system:
$Z^{ab}(E)=\case\rmi2\sum_{n}v^{a*}_nv^b_n/(E-E_{n}+\case\rmi2\Gamma)$. A
rotation that diagonalizes the (random) Hamiltonian matrix $H$ transforms
the (fixed) coupling amplitudes $V^{a}_n$ to Gaussian distributed random
coupling amplitudes $v^a_n$ with zero mean and covariances
$\aver{v^{a*}_nv^{b}_m}=(2\kappa_a\Delta/\pi)\delta^{ab}\delta_{nm}$. In
such a representation the energy correlation function acquires the following
form:
\begin{equation}\label{Zcorr}
\fl
C^{abcd}_Z(\omega)=\sum_{n,m}\frac{1}{4}\aver{v^{a}_nv^{b*}_nv^{c*}_mv^{d}_m}
\averc{ \frac{1}{E_1-E_n-\case\rmi2\Gamma}
\frac{1}{E_2-E_m+\case\rmi2\Gamma} }
\end{equation}
and the averaging over coupling amplitudes and that over the spectrum can be
done independently. The Gaussian statistics of $v$ results in
\begin{equation}\label{vvvv}
\fl \case14(\case{\pi}{\Delta})^2 \aver{v^{a}_nv^{b*}_nv^{c*}_mv^{d}_m}=
\kappa_a\kappa_c\delta^{ab}\delta^{cd} +
\kappa_a\kappa_b(\delta^{ac}\delta^{bd} +
\delta_{1\beta}\delta^{ad}\delta^{bc})\delta_{nm}
\end{equation}
where $\delta_{1\beta}$ term accounts for the presence of TRS, when all
$v^a_n$ are real and $Z$ is symmetric. It is useful to represent the
spectral correlation function in a form of the Fourier integral
$\int_{0}^{\infty}\rmd t_1\int_{0}^{\infty}\rmd t_2 e^{-\Gamma(t_1+t_2)/2}
e^{\rmi E(t_2-t_1)} e^{\rmi(E_2-E_1)(t_1+t_2)/2}
\averc{e^{\rmi(E_nt_1-E_mt_2)}}$. Due to the uniformity of local
fluctuations in the bulk of the spectrum, one can integrate additionally
over the position $E$ of the mean energy: $\int\!\frac{\rmd
E}{N\Delta}e^{\rmi E(t_2-t_1)}=\case1N\delta(\frac{t_2-t_1}{t_H})$, where
$t_{H}\equiv2\pi/\Delta$ is the Heisenberg time. It is natural therefore to
measure the time in units of $t_H$. From the known RMT spectral fluctuations
one also has for $n{\ne}m$
$(1-N)\averc{e^{2\pi\rmi(E_n-E_m)t/\Delta}}=b_{2,\beta}(t)$, where
$b_{2,\beta}(t)$ is the spectral form factor defined through the Fourier
transform of $Y_{2,\beta}(\omega)=\int_{-\infty}^{\infty}\!\rmd t
e^{2\pi\rmi\omega t}b_{2,\beta}(t)$ \cite{Mehta2,Guhr1998}:
\numparts%
\begin{eqnarray}
\fl \label{b_21}%
b_{2,\beta=1}(t)=[1-2 t+ t\log(1+2 t)]\Theta(1- t)+ [ t\log(\case{2 t+1}{2
t-1})-1]\Theta( t-1)
\\
\fl \label{b_22}%
b_{2,\beta=2}(t)= (1- t)\Theta(1- t)
\end{eqnarray}
\endnumparts%
at $ t>0$ and $b_{2,\beta}(-t)=b_{2,\beta}(t)$. Combining all these
together, we arrive finally at
\numparts%
\begin{eqnarray}
\fl \label{ZcorrFTa}%
C^{abcd}_Z(\omega)=\int_0^{\infty}\!\!\!\rmd t e^{2\pi\rmi\omega t}
\hat{C}^{abcd}_Z( t)
\\
\fl \label{ZcorrFTb}%
\hat{C}^{abcd}_Z(t)=4\,e^{-\gamma t}\,[\kappa_a\kappa_c[1-b_{2,\beta}(t)]
\delta^{ab}\delta^{cd} + \kappa_a\kappa_b(\delta^{ac}\delta^{bd} +
\delta_{1\beta}\delta^{ad}\delta^{bc}) ]\,.
\end{eqnarray}
\endnumparts%
Similar in spirit calculations were done in a context of reverberation in
complex structures in \cite{Davy1987,Lobkis2000} and in a context of chaotic
photodissociation in \cite{Alhassid1992,Alhassid1998}.

The form factor (\ref{ZcorrFTb}) is simply related  to that of $K$
matrix elements at zero absorption as $\hat{C}^{abcd}_Z(
t)=e^{-\gamma t} \hat{C}^{abcd}_K( t)$. Such a relationship
between the corresponding form factors with and without absorption
is generally valid for any correlation function which may be
reduced to the two-point correlation function of resolvents (see
\cite{Schaefer2003} and the discussion below, e.g., for the case
of the $S$ matrix). This can be easily understood as a result of
the analytic continuation $2\pi\omega\to2\pi\omega+i\gamma$ of the
energy difference $\omega$ reflecting switching on the absorption
(see the previous section).

The obtained expressions describe a gradual loss of correlations
in $Z$ matrix elements as the energy difference grows; generally,
$C_Z(\omega{\to}\infty)\to0$. At $\omega=0$, (\ref{ZcorrFTa})
provides us with impedance variances
$C_Z^{abab}(0)=\mathrm{var}(Z^{ab})\equiv
\langle|Z^{ab}|^2\rangle-|\langle{Z^{ab}}\rangle|^2$, which were
recently studied experimentally in \cite{Zheng2005}. In analogy
with the so-called elastic enhancement factor considered
frequently in nuclear physics \cite{Verbaarschot1986}, one can
define the following ratio of variances in reflection ($a{=}b$) to
that in transmission ($a{\neq}b$):
\begin{equation}\label{W_Z}
W_{Z,\beta} \equiv \frac{\sqrt{\mathrm{var}(Z^{aa})\mathrm{var}(Z^{bb})} }{
\mathrm{var}(Z^{ab})} = 2+\delta_{1\beta} - \int_{0}^{\infty}\!\!\rmd s\,
e^{-s} b_{2,\beta}(\case{s}{\gamma})
\end{equation}
where the second equality follows easily from (\ref{ZcorrFTb}) (note that
the coupling constants $\kappa_{a,b}$ are mutually cancelled here). Making
use of $b_{2,\beta}(\infty){=}0$ and $b_{2,\beta}(0){=}1$, one can readily
find $W_{Z,\beta}$ in the limiting cases of weak or strong absorption as:
\begin{equation}\label{W_Zlim}
W_{Z,\beta} =\cases{2+\delta_{1\beta} & at $\gamma\ll1$\\
1+\delta_{1\beta} & at $\gamma\gg1$\\} \,.
\end{equation}
$W_{Z,\beta}$ decays monotonically as absorption grows. In the
case of unitary symmetry, (\ref{b_22}) and (\ref{W_Z}) yield
explicitly $W_{Z,2}=1+\case1\gamma(1-e^{-\gamma})$ in agreement
with \cite{Zheng2005}. It is hardly possible to get a simple
explicit expression at finite $\gamma$ in the case of orthogonal
symmetry. However, a reasonable approximation can be found if one
notices that the integration in (\ref{W_Z}) is determined mainly
by the region $s\leq1$, so that one can approximate
$b_{2,1}(s)\approx(1-2s+2s^2)\Theta(1-s)$ through its Taylor
expansion. Performing the integration, one arrives at
$W_{Z,1}\approx3-\gamma^{-2}[(4+\gamma^2)(1-e^{-\gamma})
-2\gamma(1+e^{-\gamma})]$, which turns out to be a good
approximation to the exact answer at moderately strong absorption
(deviations are seen numerically only at $\gamma\sim1$).

\subsection{Scattering matrix}

The energy correlation function of the scattering matrix elements
\begin{equation}\label{Scorr}
C^{abcd}_S(\omega)\equiv\averc{S_{\gamma}^{ab*}(E_1)S_{\gamma}^{cd}(E_2)}
=\int_0^{\infty}\!\!\rmd t e^{2\pi\rmi\omega t} \hat{C}^{abcd}_S(t)
\end{equation}
is a much more complicated object for an analytical consideration
as compared to (\ref{Zcorr}). The reason becomes clear if one
considers again the pole representation of the $S$ matrix which
follows from (\ref{Heff}):
$S^{ab}(E)=\delta^{ab}-\rmi\sum{w^{a}_n\tilde{w}^{b}_n}/(E-\mathcal{E}_n)$.
Due to unitarity constraints imposed on $S$ at real $E$ the
residues and complex resonance energies develop nontrivial
correlations \cite{Sokolov1989}. Although results on statistics of
resonances in the complex plane
\cite{Fyodorov1997,Sommers1999,Fyodorov1999,Fyodorov2003} as well
as the corresponding residues \cite{Fyodorov2003,Fyodorov2002ii}
became recently available in some particular cases, this knowledge
is insufficient as yet for calculating $S$-matrix correlations in
general. The separation like (\ref{Zcorr}) into a ``coupling'' and
``spectral'' average is no longer possible and can be done only by
involving some approximations \cite{Gorin2002}. The powerful
supersymmetry method \cite{Efetov1983,Verbaarschot1985} turns out
to be an appropriate technique to perform the statistical average
in this case. In their seminal paper \cite{Verbaarschot1985},
Verbaarschot, Weidenm\"uller and Zirnbauer performed the exact
calculation of (\ref{Scorr}) at arbitrary transmission
coefficients (and zero absorption) in the case of orthogonal
symmetry. This finding was later adopted \cite{Schaefer2003} to
include absorption. The corresponding exact result for unitary
symmetry has been recently presented by us in \cite{Savin2005i}
(see also \cite{Pluhar1995} concerning the $S$-matrix variance in
the GOE-GUE crossover at perfect coupling) and is discussed below.

The calculation proceeds along the same lines as in
\cite{Verbaarschot1985}. The final expression for both the
connected correlation function and its form factor (\ref{Scorr})
can be equally represented as follows:
\begin{eqnarray}\label{ScorrFT}
\fl C^{abcd}_S= \delta^{ab}\delta^{cd}T_aT_c\sqrt{(1-T_a)(1-T_c)} J_{ac} +
(\delta^{ac}\delta^{bd}+\delta_{1\beta}\delta^{ad}\delta^{bc})T_aT_bP_{ab}\,.
\end{eqnarray}
Here, the $\delta_{1\beta}$ term accounts trivially for the symmetry
property $S^{ab}=S^{ba}$ in the presence of TRS. $J_{ac}$ and $P_{ab}$
defined below are some functions (of the energy difference $\omega$ or the
time $t$), which depend also on TRS, coupling and absorption but already in
a nontrivial way. As a result, the elastic enhancement factor
$W_{S,\beta}\equiv\sqrt{\mathrm{var}(S^{aa})\mathrm{var}(S^{bb})}/\mathrm{var}(S^{ab})$
is generally a complicated function of all these parameters, in contrast to
(\ref{W_Z}). In the particular case of perfect coupling, all $T_{c}=1$, one
has obviously from (\ref{ScorrFT}) that $W_{S,\beta}=1+\delta_{1\beta}$ at
any absorption strength.

We consider first expression (\ref{ScorrFT}) in the energy domain at real
$\omega$ (no absorption). The functions $J_{ac}(\omega)$ and
$P_{ab}(\omega)$  can be generally written as certain expectation values in
the field theory (nonlinear zero-dimensional supersymmetric $\sigma$-model)
whose explicit representations depend on the symmetry case considered, we
refer the reader to \cite{Verbaarschot1985,Efetov} for general discussion.
In the $\beta=1$ case of orthogonal symmetry the well-known result of
\cite{Verbaarschot1985} reads as follows:
\numparts %
\begin{eqnarray}
\fl J_{ac}(\omega) = \textstyle \aver{
\bigl(\frac{\mu_1}{1+T_a\mu_1}+\frac{\mu_2}{1+T_a\mu_2}+\frac{\mu_0}{1-T_a\mu_0}\bigr)
\bigl(\frac{\mu_1}{1+T_c\mu_1}+\frac{\mu_2}{1+T_c\mu_2}+\frac{\mu_0}{1-T_c\mu_0}\bigr)
\mathcal{F}_M}_{\mu}
\\ \textstyle
\fl P_{ab}(\omega) = \textstyle \aver{ \bigl(
\frac{\mu_1(1+\mu_1)}{(1+T_a\mu_1)(1+T_b\mu_1)}
+\frac{\mu_2(1+\mu_2)}{(1+T_a\mu_2)(1+T_b\mu_2)}
+\frac{\mu_0(1-\mu_0)}{(1-T_a\mu_0)(1-T_b\mu_0)} \bigr) \mathcal{F}_M}_{\mu}
\end{eqnarray}
\endnumparts
with
$\mathcal{F}^{}_{M}=\prod_c[\frac{(1-T_c\mu_0)^2}{(1+T_c\mu_1)(1+T_c\mu_2)}]^{1/2}$
being the so-called ``channel factor'', which accounts for system openness,
and $\langle(\cdots)\rangle_{\mu}$ is to be understood explicitly as
\begin{equation}\label{mu}
\fl \frac{1}{8} \int_{0}^{\infty}\!\rmd\mu_1 \!\int_{0}^{\infty}\!\rmd\mu_2
\!\int_{0}^{1}\!\rmd\mu_0 \frac{(1-\mu_0)\mu_0
|\mu_1-\mu_2|\,e^{\rmi\pi\omega(\mu_1+\mu_2+2\mu_0)} }{
[(1+\mu_1)\mu_1(1+\mu_2)\mu_2]^{1/2} (\mu_0+\mu_1)^2 (\mu_0+\mu_2)^2 }
\left(\ldots \right)\,.
\end{equation}
In the $\beta=2$ case of unitary symmetry we have found \cite{Savin2005i}
that
\numparts %
\begin{eqnarray}
J_{ac}(\omega) = \textstyle \aver{
\bigl(\frac{\mu_1}{1+T_a\mu_1}+\frac{\mu_0}{1-T_a\mu_0}\bigr)
\bigl(\frac{\mu_1}{1+T_c\mu_1}+\frac{\mu_0}{1-T_c\mu_0}\bigr)
\mathcal{F}_M}_{\mu}
\\ \textstyle
P_{ab}(\omega) = \textstyle \aver{ \bigl(
 \frac{\mu_1(1+\mu_1)}{(1+T_a\mu_1)(1+T_b\mu_1)}
+\frac{\mu_0(1-\mu_0)}{(1-T_a\mu_0)(1-T_b\mu_0)} \bigr) \mathcal{F}_M}_{\mu}
\end{eqnarray}
\endnumparts
with the channel-factor
$\mathcal{F}_{M}=\prod_c\frac{1-T_c\mu_0}{1+T_c\mu_1}$ and the corresponding
integration being
\begin{equation}
\int_{0}^{\infty}\!\rmd\mu_1\int_{0}^{1}\!\rmd\mu_0
(\mu_1+\mu_0)^{-2}\,e^{\rmi2\pi\omega(\mu_1+\mu_0)} \left(\ldots \right)\,.
\end{equation}
In the important particular case of the single open channel (elastic
scattering), the general expression for the $\beta=2$ case simplifies
further to
\begin{equation}
\fl \averc{S^*(E_1)S(E_2)}=T^2\int_{0}^{\infty}\!\rmd\mu_1\int_{0}^{1}
\!\frac{\rmd\mu_0}{\mu_1+\mu_0}
\frac{1+(2-T)\mu_1}{(1+T\mu_1)^3}\,e^{\rmi2\pi\omega(\mu_1+\mu_0)}\,.
\end{equation}
Finally, putting above $\omega\to\omega+\rmi\gamma/2\pi$ accounts for the
finite absorption strength $\gamma$.

To consider (\ref{ScorrFT}) in the time domain, i.e. the form factor
$\hat{C}^{abcd}_S(t)$, we notice that the variable $
t=\case12(\mu_1+\mu_2+2\mu_0)$ for $\beta=1$ or $ t=\mu_1+\mu_0$ for
$\beta=2$ plays the role of the dimensionless time (in units of $t_H$). The
corresponding expressions for $\hat{P}_{ab}(t)$ and $\hat{J}_{ac}(t)$ can be
investigated using the methods developed in
\cite{Verbaarschot1986,Gorin2002,Dittes2000}. For orthogonal symmetry it was
done in \cite{Schaefer2003}, where the overall decaying factor $e^{-\gamma
t}$ due to absorption was also confirmed by comparison to the experimental
result for the form factor measured in microwave cavities. It is useful for
the qualitative description to note that $\hat{P}_{ab}(t)$ and
$2\hat{J}_{ac}(t)$ are quite similar to the ``norm leakage'' decay function
\cite{Savin1997} and the form factor of the Wigner's time delays
\cite{Lehmann1995b}, respectively (they would coincide exactly at
$\gamma=0$, if we put $T_{a,b,c}=0$ appearing explicitly in denominators
above). Then one can follow analysis performed there, see also
\cite{Dittes2000}, to find $\hat{P}_{ab}(t)\approx e^{-\gamma t}$ and
$\hat{J}_{ac}(t)\approx(2t/\beta)e^{-\gamma t}$ as exact asymptotic at small
times \cite{Gorin2002}, while they both become proportional to $e^{-\gamma
t} t^{-M\beta/2-2}$ at large times.

Such a power law  is characteristic for open systems
\cite{Lewenkopf1991,Dittes2000,Savin1997}. Physically, it results from width
fluctuations, which diminish as the number $M$ of open channels grows
\cite{Fyodorov1997,Savin1997}. In the limiting case $M\to\infty$ and
$T_{c}\to0$, all the resonances acquire just the same escape width
$\sum_cT_{c}$ (in units of $t_{H}^{-1}$), which is often called the
Weisskopf's width \cite{Blatt}, so that the total width is
$\gamma_T=\sum_cT_{c}+\gamma$. Then there occur further simplifications:
$\hat{P}_{ab}( t)=e^{-\gamma_T t}$ and $\hat{J}_{ab}( t)=[1-b_{2,\beta}(
t)]e^{-\gamma_T t}$, that results finally in
\begin{equation}\label{ScorrMT}
\fl C^{abcd}_S(\omega) =
\frac{(\delta^{ac}\delta^{bd}+\delta_{1\beta}\delta^{ad}\delta^{bc})T_aT_b
}{ \gamma_T-2\pi\rmi\omega } +
\delta^{ab}\delta^{cd}T_aT_c\int_{0}^{\infty}\!\rmd t [1-b_{2,\beta}(
t)]e^{-(\gamma_T-2\pi\rmi\omega) t}.
\end{equation}
For the case of $\beta=1$ this result (at zero absorption) was obtained
earlier by Verbaarschot \cite{Verbaarschot1986}. In the limit considered,
expression (\ref{ScorrMT}) is very similar to (\ref{ZcorrFTa}),
(\ref{ZcorrFTb}), so that the enhancement factor $W_{S,\beta}$ is given by
the same (\ref{W_Z}) where $\gamma$ is to be substituted with $\gamma_{T}$.
At $\gamma_T\gg1$ (large resonance overlapping or strong absorption, or
both) the dominating term in (\ref{ScorrMT}) is the first one, which is
known as the Hauser-Feshbash relation \cite{Hauser1952}, see
\cite{Moldauer1975a,Agassi1975,Friedman1985} for discussion. Then
$W_{S,\beta}=2/\beta=W_{Z,\beta}$ that can be also understood as the
consequence of the gaussian statistics of $S$ (as well as of $Z$) in the
limit of strong absorption \cite{Friedman1985}.

\subsection{Reflection}

In this subsection we consider fluctuations in the weighted-mean
reflection coefficient defined as $r=\case1M\tr R$. Its average
value $\aver{r}$ was recently calculated in \cite{Savin2003i},
where the following exact result was found to be
\begin{equation}\label{<r>}
\langle{r}\rangle=1-\frac{\gamma}{M}\left[1-\gamma\int_0^{\infty}\!\!\rmd
t\, e^{-\gamma t}P( t)\right]
\end{equation}
relating $\aver{r}$ to the so-called ``norm-leakage'' function $P( t)$
introduced in \cite{Savin1997}. The actual value of $\aver{r}$ changes with
growing absorption between $1-\frac{\gamma}{M}$ at $\gamma\ll1$ and
$1-\frac{1}{M}\sum_{c}T_c$ at $\gamma\gg M$. For the sake of simplicity we
consider the case of $M$ equivalent channels below, all $T_c=T$. For large
$M\gg1$ the widths cease to fluctuate, yielding $P( t)=e^{-MT t}$, so that
(\ref{<r>}) results in $\aver{r}=(MT+\gamma(1-T))/(MT+\gamma)$
\cite{Savin2004}.

The correlation function
$C_{R}(\omega)=\averc{r(E_1)r(E_2}=\Gamma^2\averc{Q_w(E_1)Q_w(E_2}$
is simply related by virtue of (\ref{R}) to that of Wigner's time
delay $Q_w=\case1M\tr{Q_{\gamma}}$ (at finite $\Gamma$). Amounting
to a four-point correlation function of resolvents, its evaluation
is generally beyond the present day state of art in the
supersymmetry method. However, by approximating
$Q_w\approx\frac{1}{M}\sum_n\Gamma_n/[(E-E_n)^2+\case14(\Gamma_n+\Gamma)^2]$,
cf. (\ref{R}), and making use of the rescaled Breit-Wigner
approximation (RBWA) of Gorin and Seligman \cite{Gorin2002}, one
can find an expression which is supposed \cite{Schaefer2003} to
work reasonably well at finite absorption. The final result can be
naturally represented in the form of the Fourier-integral
\begin{equation}\label{C_R}
C_{R}(\omega) = \frac{2\gamma^2}{M^2} \int_{0}^{\infty}\!\rmd
t\cos(2\pi\omega t)\,e^{-\gamma t}F_R(t)
\end{equation}
where the form-factor (we have set apart in (\ref{C_R}) the trivial
absorbing factor $e^{-\gamma t}$) is
\begin{equation}\label{F_R}
F_R(t) = \frac{c_{2,\beta}( t)}{(1+2T t/\beta)^{M\beta/2}} -
\frac{b_{2,\beta}( t)\,c_{1,\beta}(\case t2)^2}{(1+Tt/\beta)^{M\beta}}
\end{equation}
with the function $c_{n,\beta}( t)$, $n=1,2$, defined as follows:
\begin{equation}
c_{n,\beta}(t) = \prod_{l=0}^{n-1}(\case12M\beta+l) \int_{0}^{\infty}\!\rmd
x \frac{x^{n-1}\,e^{-(\beta\gamma/2T)(1+2Tt/\beta)x}}{ (1+x)^{M\beta/2+n}}
\,.
\end{equation}

It is instructive to consider the limiting cases of weak and strong
absorption in more detail. At $\gamma\ll T \le1$, the $\gamma$-dependence
and thus $t$-dependence of $c_{n,\beta}(t)$ is very weak, so that
$c_{n,\beta}(t)\approx1$ for $t\ll1/\gamma$. As a result, the form-factor
reduces to
\begin{equation}\label{Fweak}
F_R(t) = (1+2Tt/\beta)^{-M\beta/2} - b_{2,\beta}(t)(1+Tt/\beta)^{-M\beta}
\end{equation}
which is, as expected, just the form-factor of the Wigner's
time-delay correlation function at \emph{zero} absorption
calculated in the same scheme RBWA. In the opposite limiting case
$\gamma\gg T$, the $x$-integration in $c_{n,\beta}(t)$ is
determined mainly by the region of small $x\le{T/\gamma}\ll1$,
that yields readily
\begin{equation}\label{Fstrong}
F_R(t) = \left(\frac{MT}{\gamma}\right)^2
\left[\frac{1+2/M\beta}{(1+2Tt/\beta)^{M\beta/2+2}} -
\frac{b_{2,\beta}(t)}{(1+Tt/\beta)^{M\beta+2}}\right]\,.
\end{equation}
Comparing the large time asymptotic of $F_R(t)$, one sees that a
\textit{crossover} from $t^{-M\beta/2}$ to $t^{-M\beta/2-2}$ behavior
happens at $\gamma\sim T$, as we go from weak to strong absorption regime.
This behaviour is indicative of an additional decay of correlations induced
by strong absorption on top of the pure exponential one. It is most
pronounced for a single-channel cavity with TRS, corresponding to changing
from $t^{-1/2}$ to $t^{-5/2}$.

\section{Distribution functions}

\subsection{Single-channel scattering}
\subsubsection{Reflection coefficient and the local Green function.}

It is quite clear, that the joint distribution of the real and imaginary
parts of $K$
\begin{equation}\label{P(u,v)def}
\mathcal{P}(u,v) = \left\langle
\delta(u-\mathrm{Re\,}K)\,\delta(v+\mathrm{Im\,}K)\right\rangle
\end{equation}
determines fully the statistics of the impedance and the
$S$-matrix. By definition (\ref{S}), $K=\kappa(N\Delta/\pi)G_{ii}$
is related to the diagonal element
$G_{ii}=\langle{i}|(E_\gamma-H)^{-1}|i\rangle$ of the Green
function  of the closed system in the position representation,
taken at the position $i$ of channel attachment. The RMT
assumption for $H$ implies however that the  basis of
representation may be chosen arbitrary. One studies, therefore,
statistics of the dimensionless local Green function $K\equiv
u-iv$, with $v>0$ being the local density of states (LDoS). We
consider first the case of \emph{perfect} coupling $\kappa=1$
($T=1$). In particular, this choice ensures the normalisation of
the mean LDoS such that $\langle{v}\rangle=1$.

In this case, the following general form of the distribution function
\begin{equation}\label{P(u,v)}
{\mathcal{P}}(u,v) = \frac{1}{2\pi v^2}P_0(x)\,,\qquad
x=\frac{u^2+v^2+1}{2v}>1.
\end{equation}
may be easily established \cite{Fyodorov2004ii}. It initially emerged in
\cite{Mirlin1994i,Mirlin1994ii} in the course of tedious calculations for
the $\beta{=}2$ symmetry class, but neither origin nor generality of such a
form were appreciated. Actually, the validity of the form (\ref{P(u,v)})
follows directly from exploiting the two following fundamental properties of
the $S$-matrix statistics at perfect coupling: (i) the uniform distribution
of the scattering phase $\theta\in(0,2\pi)$; and (ii) the statistical
independence of $\theta$ and the $S$ matrix modulus. In the cases of chaotic
systems with pure symmetries, both these properties can be verified making
use of methods of \cite{Brouwer1997ii}. In this paper we are proving
(\ref{P(u,v)}) directly by an alternative, more powerful method
\cite{Savin2005}. It holds much beyond the universal RMT regime for a very
broad class of disordered Hamiltonians, in particular for the case when
localisation effects already play an important role. Moreover, this method
will help us to verify that (\ref{P(u,v)}) holds in the crossover regime
between the pure ensembles. For the time being the uniformity of the
scattering phase distribution will be taken for granted.

Substituting $K=u-\rmi v$ into (\ref{S}), we can immediately infer that the
variable $x$ is directly related to the reflection coefficient,
parameterizing it in the following way:
\begin{equation}\label{r-x}
S_{\gamma}\equiv\sqrt{r}e^{\rmi\theta}\,, \qquad r=\frac{x-1}{x+1}\,, \qquad
\cot\theta=\frac{u^2+v^2-1}{2u}\,.
\end{equation}
The joint distribution $\mathcal{P}(u,v)$ of $u$ and $v$ and that
$P(x,\theta)$ of $x$ and $\theta$ are related by means of the
Jacobian $|\frac{\partial(u,v)}{\partial(x,\theta)}|=v^{-2}$ as
can be verified by a straightforward calculation. Finally,
$P(x,\theta)=\frac{1}{2\pi}P_0(x)$ results from the uniform
distribution of the phase and in this way $P_0(x)$ acquires the
physical meaning of the normalized distribution of the reflection
parameter $x$.

Relationship (\ref{P(u,v)}) is one of our central results, which despite its
apparent simplicity will have many far-reaching consequences which are not
easy to guess otherwise.  Let us note the invariance property of this
distribution under the change $iK\to1/iK$, meaning that both the impedance
and its inverse (i.e. the admittance) must have one and the same probability
distribution. Another remarkable feature of (\ref{P(u,v)}) is, that it
relates the distribution of the local Green function in the completely
\emph{closed} system to the distribution of the reflection coefficient in
the perfectly \emph{open} one. This fact will be fully utilised later on for
extracting statistical properties of the Wigner time delay.

It is instructive to discuss first the \emph{exact} limiting statistics
\begin{equation}\label{P(x)lim}
\fl P_0(x) = \left\{ \begin{array}{ll}
\frac{\alpha^{\beta/2+1}}{2\Gamma(\beta/2+1)}
\left(\frac{x+1}{2}\right)^{\beta/2}e^{-\alpha(x+1)/2} &
(\alpha\ll1) \\[2ex]
\case\alpha2\,e^{-\alpha(x-1)/2} & (\alpha\gg 1)
\end{array}\right.\,, \qquad \alpha\equiv\case12\beta\gamma\,,
\end{equation}%
which can be found \cite{Fyodorov2004ii} in the weak or strong absorption
limit. In the former case, one can follow
\cite{Ramakrishna2000,Beenakker2001} in using a perturbation theory to
relate reflection to the time delay in the ideal cavity without absorption
(see (\ref{R}) thereafter). The latter quantity has a known distribution
\cite{Fyodorov1997,Gopar1996}
$\mathcal{P}_{\tau}(\tau=\case{q}{t_H})=[(\case\beta2)^{\beta/2}/\Gamma(\case\beta2)]
\tau^{-\beta/2-2}e^{-\beta/2\tau}$ , which yields the first line above. In
the opposite case of strong absorption, the real and imaginary parts of
$S$-matrix become statistically independent Gaussian distributed variables
\cite{Friedman1985,Kogan2000}, resulting in the so-called Rayleigh
distribution $P(r)\simeq(\gamma\beta/2)e^{-r\gamma\beta/2}$ of reflection
valid at $r\approx\case12(x-1)\ll1$, and the second line in (\ref{P(x)lim})
follows.

As to arbitrary absorption, an exact result was obtained first for
unitary symmetry ($\beta=2$) by Beenakker and Brouwer
\cite{Beenakker2001}. It is convenient to keep the scaled
absorption parameter $\alpha$, representing their result in the
following form
\begin{equation}\label{P(x)gue}
P_0(x) = \frac{{\mathcal N}_{\beta}}{2}
\left[A\,\left(\alpha(x+1)/2\right)^{\beta/2}+B\right]e^{-\alpha(x+1)/2}
\end{equation}
with $\alpha$-dependent constants $A\equiv e^{\alpha}-1$ and
$B\equiv1+\alpha-e^{\alpha}$, and with $\mathcal{N}_2=1$ standing
for the normalization. For the case of symplectic symmetry
($\beta=4$), the exact form was reported very recently
\cite{Fyodorov2004ii}. The explicit derivation is given in
Appendix A, the final result being:
\begin{equation}\label{P(x)gse}
\fl P_0^{\mathrm{gse}}(x) = \widetilde{P}_0^{\mathrm{gue}}(x) +
\Bigl[\frac{\gamma^2}{2}(x+1)^2-\gamma(\gamma+1)(x+1)+\gamma\Bigr]\,
e^{-\gamma x}\! \int_{0}^{\gamma}\!\!\rmd t\frac{\sinh{t}}{t}
\end{equation}
where $\widetilde{P}_0^{\mathrm{gue}}(x)$ is the distribution
(\ref{P(x)gue}) for $\beta=2$ taken, however, at $\alpha=2\gamma$. The case
of orthogonal symmetry, the most welcomed experimentally, turns out to be
extremely tricky. It was suggested in \cite{Fyodorov2004ii}, see also
\cite{Kuhl2005}, that expression (\ref{P(x)gue}) (with
$\mathcal{N}_{\beta}=\alpha/(A\Gamma(\beta/2{+}1,\alpha)+Be^{-\alpha})$ and
$\Gamma(\nu,\alpha)=\int_{\alpha}^{\infty}\!\rmd p\,p^{\nu-1}e^{-p}$) is an
appropriate interpolation formula at $\beta=1$. It incorporates correctly
both known limiting cases of weak or strong absorption and a reasonable
agreement with available numerical and experimental data was found there in
a broad range of the absorption strength. Below we provide an exact
analytical treatment of this case by the method suggested in
\cite{Savin2005}.

\subsubsection{$\mathcal{P}(u,v)$ and the spectral correlation function of Green's
function resolvents.}

We establish now the general relation \cite{Savin2005} between the joint
distribution function (\ref{P(u,v)}) at arbitrary \emph{finite} absorption
and the energy autocorrelation function
\begin{equation}\label{C_omega}
\fl C_{\Omega}(z_-,z_+) = \left\langle
\frac{1}{z_{-}-\rmi0-K_{0}(E-\case12\Omega-\rmi0)}
\frac{1}{z_{+}+\rmi0-K_{0}(E+\case12\Omega+\rmi0)} \right\rangle
\end{equation}
of the resolvents of the local Green function $K_0$ at zero absorption
($\Gamma=0$, indicated explicitly in this subsection with the subscript
``0''). Distribution (\ref{P(u,v)}) can be obtained from (\ref{C_omega}) by
analytic continuation in $\Omega$ from a real to purely imaginary value
$\Omega=\rmi\Gamma$ as follows. $K_0(E)$ is an analytic function of the
energy in the upper or lower half-plane and, thus, can be  analytically
continued to the complex values: $K_0(E\pm\case\rmi2\Gamma)\equiv u\mp\rmi
v$, $v>0$. This allows us to continue then analytically the correlation
function (\ref{C_omega}) from a pair of its real arguments to the complex
conjugate one: $z_+ = (z_-)^*\equiv z'+\rmi z''$, $z''>0$. As a result,
function (\ref{C_omega}) acquires at $\Omega=\rmi\Gamma$ the following form:
\begin{eqnarray}\label{C}
\fl C(z',z'') \equiv C_{\Omega=i\Gamma}(z_-,z_+) =
    \left\langle\frac{1}{(z'-u)^2+(z''+v)^2}\right\rangle \nonumber\\
= \int_{-\infty}^{\infty}\!\rmd u\!\int_{0}^{\infty}\!\rmd v\,
    \frac{\mathcal{P}(u,v)}{(z'-u)^2+(z''+v)^2} \,.
\end{eqnarray}
The second line here is due to the definition (\ref{P(u,v)}). To solve this
equation for $\mathcal{P}(u,v)$, we perform first the Fourier transform (FT)
with respect to $z'$ that leads to
\begin{equation}\label{Chat}
\fl \widehat{C}(k,z'') \equiv \int_{-\infty}^{\infty}\!\rmd
z'e^{ikz'}C(z',z'') = \int_{0}^{\infty}\!\rmd
v\,\widehat{\mathcal{P}}(k,v)\, \frac{\pi\,e^{-|k|(z''+v)}}{z''+v}\
\end{equation}
where $\widehat{\mathcal{P}}(k,v)$ is the corresponding FT of
$\mathcal{P}(u,v)$. Being derived at $z''>0$, Eq.~(\ref{Chat}) can be
analytically continued to the whole complex $z''$ plane with a cut along
negative $\mathrm{Re\,}z''$. Calculating then the jump of
$\widehat{C}(k,z'')$ on the discontinuity line $z''=-v$ ($v>0$), we finally
get the following expression
\begin{equation}\label{Phat}
\widehat{\mathcal{P}}(k,v)
=\frac{\Theta(v)}{2\pi^2\rmi}\left[\widehat{C}(k,-v-\rmi0) -
\widehat{C}(k,-v+\rmi0)\right]
\end{equation}
with the Heaviside step function $\Theta(v)$. The inverse FT of (\ref{Phat})
yields $\mathcal{P}(u,v)$.

This completely general relationship is another our central
result. It resembles (and reduces to) the well-known relation
between the spectral density of states and the imaginary part of
the corresponding resolvent operator, when the case of one real
variable is considered. In contrast, the case of the distribution
of two real variables requires to deal with the two-point
correlation function. Physically, the latter is a generalized
susceptibility describing a response of the system under
consideration. This fact suggests to consider our formula in a
sense as a ``fluctuation dissipation'' relation: The l.h.s. of
(\ref{Phat}) stands for the \emph{distribution} (of $K$) in the
presence of dissipation / absorption whereas the \emph{correlation
function} (of resolvents of $K$) in the r.h.s. accounts for
fluctuations in the system, i.e. for arbitrary order correlations
in the absence of absorption.

The main advantage of the derived relation is that the correlation function
is a much easier object to calculate analytically as compared to the
distribution. Such a calculation for ideal systems at zero absorption has
actually already been performed in many interesting cases. In the particular
case of a chaotic cavity an exact result for the correlation function
(\ref{C_omega}) has been previously derived by us in
\cite{Fyodorov1997,Fyodorov1997i}. Moreover, similar formulae can be derived
for any model describing a one-particle quantum motion in a d-dimensional
sample with a static disordered potential, see outline of the derivation in
Appendix B. The only important physical assumption is that the disorder is
such that all the relevant statistical properties of the system can be
adequately described by the standard diffusive supersymmetric non-linear
$\sigma$-model (or its lattice version). For a detailed discussion of the
validity of this approximation (and its limitations) the interested reader
is referred to the review \cite{Mirlin2000} and the book \cite{Efetov}.

In the ``zero-dimensional'' case (chaotic cavity) the analytic continuation
of (\ref{C_omega}) to complex $\Omega=\rmi\Gamma$ can be represented
generally as follows (see Appendix B for details):
\begin{equation}\label{Cgeneric}
\fl C(z',z'') = \frac{1}{z'^2+(z''+1)^2} + \frac{1}{4}\left(
\frac{\partial^2}{\partial z'^2} \!+\! \frac{\partial^2}{\partial z''^2}
\right) \mathcal{F}(\tilde{x})\,,\qquad
\tilde{x}\equiv\frac{z'^2+z''^2+1}{2z''}
\end{equation}
where it is important that the function $\mathcal{F}(\tilde{x})$ depends on
$z'$ and $z''$ only via the scaling variable $\tilde{x}>1$. Its explicit
form depends on the symmetry present (e.g. preserved or broken TRS), the
following common structure being however generic:
\begin{equation}\label{Fgeneric}
\fl \mathcal{F}(\tilde{x}) =
\int_{-1}^{1}\!\rmd\lambda_0\!\int_{1}^{\infty}\!\rmd\lambda_1
\int_{1}^{\infty}\!\rmd\lambda_2 \frac{ \,f(\{\lambda\})\,
e^{-\gamma(\lambda_1\lambda_2-\lambda_0)/2}\, (\tilde{x}+\lambda_0) }{
[(\tilde{x}+\lambda_1\lambda_2)^2- (\lambda_1^2-1)(\lambda_2^2-1)]^{1/2}}\,.
\end{equation}
The \emph{real} function $f(\{\lambda\})$ is the only symmetry dependent
term here. In the crossover regime of gradually broken TRS it can be
represented explicitly as follows \cite{Fyodorov1997i}:
\begin{eqnarray}\label{f}
\fl f(\{\lambda\}) = \left\{ (1-\lambda_0^2)(1+e^{-2Y}) \right.
-(\lambda_1^2-\lambda_2^2)(1-e^{-2Y}) \nonumber\\ \lo + 4y^2\mathcal{R}[
\left.(1-\lambda_0^2)e^{-2Y} + \lambda_2^2(1-e^{-2Y})] \right\}
\frac{e^{-2y^2(\lambda_2^2-1)}}{\mathcal{R}^2}
\end{eqnarray}
with
$\mathcal{R}=\lambda_0^2+\lambda_1^2+\lambda_2^2-2\lambda_0\lambda_1\lambda_2-1$
and $Y\equiv{y^2}(1-\lambda_0^2)$, where $y$ denotes a crossover driving
parameter. Physically, $y^2\sim{\delta}E_y/\Delta$ is determined by the
energy shift $\delta{E}_y$ of energy levels due to a TRS breaking
perturbation (e.g., weak external magnetic field in the case of quantum
dots) \cite{Dupius1991,Altland1992}. Such an effect is conventionally
modelled within the framework of RMT by means of the ``Pandey-Mehta''
Hamiltonian \cite{Pandey1983}, $H=\hat{H}_{S}+\rmi(y/\sqrt{N})\hat{H}_{A}$,
with $\hat{H}_{S}$ ($\hat{H}_{A}$) being a random real symmetric
(antisymmetric) matrix with independent Gaussian distributed entries. The
limit $y\to0$ or $y\to\infty$ corresponds to fully preserved or broken TRS,
respectively.

Now we apply relation (\ref{Phat}) to Eq.~(\ref{Cgeneric}) and then perform
the inverse FT to get $\mathcal{P}(u,v)$. Relegating all technical details
to Appendix C, we emphasize here the most important points. The nontrivial
contribution to the distribution comes from the second (``connected'') part
of the correlation function (\ref{Cgeneric}) whereas the first
(``disconnected'') one is easily found to yield the singular contribution
$\delta(u)\delta(v-1)$. A careful analysis shows that due to specific
$\tilde{x}$-dependence given by Eq.~(\ref{Fgeneric}) the above described
procedure for the analytic continuation of the connected part of
$\widehat{C}(k,z'')$ is equivalent to continuing $\mathcal{F}(\tilde{x})$
analytically and taking the jump at
\begin{equation}\label{continuation}
\tilde{x}=-x\pm{i}0,\quad \mathrm{with}\quad
x\equiv\case1{2v}(u^2+v^2+1)>1\,.
\end{equation}
Thus, the nonzero imaginary part $F(x)=\mathrm{Im}\mathcal{F}(-x+\rmi0)$ of
the analytic continuation of (\ref{Fgeneric}) is determined at given $x$ by
the following integral
\begin{eqnarray}\label{Fcross}
\fl F(x) =
\int_{-1}^{1}\!\rmd\lambda_0\!\int\!\!\int_{\mathcal{B}_x}\!\rmd\lambda_1\rmd\lambda_2
\frac{ f(\{\lambda\})\, e^{-\gamma(\lambda_1\lambda_2-\lambda_0)/2}
\,(x-\lambda_0)}{
[(\lambda_1^2-1)(\lambda_2^2-1)-(\lambda_1\lambda_2-x)^2]^{1/2}}
\end{eqnarray}
over the integration region $\mathcal{B}_x=\{(\lambda_1,\lambda_2)|$
$1\leq\lambda_{1,2}<\infty$,
$(\lambda_1\lambda_2-x)^2<(\lambda_1^2-1)(\lambda_2^2-1)\}$, where the
square root in (\ref{Fgeneric}) attains  pure imaginary values. Taking now
into account the following identity $ \left(\frac{\partial^2}{\partial
u^2}+\frac{\partial^2}{\partial v^2}\right) F(x) = v^{-2}\frac{\rmd}{\rmd
x}(x^2-1)\frac{\rmd}{\rmd x}F(x)$, which is valid for $x^2\neq1$, we arrive
finally at
\begin{equation}\label{Pgeneric}
\mathcal{P}(u,v) = \frac{1}{4\pi^2v^2}\frac{\rmd}{\rmd x}(x^2-1)\frac{\rmd
F(x)}{\rmd x} \equiv \frac{1}{2\pi v^2}P_0(x)\,.
\end{equation}
This proves in general, cf. (\ref{P(u,v)}), the statistical independence and
uniform distribution of the scattering phase  at perfect coupling. At
arbitrary values of the crossover parameter $y$ the obtained expression can
be further treated only numerically. However, further analytical progress is
possible in the limiting cases of pure symmetries and is considered below.
We discuss explicitly the two most important cases of unitary and orthogonal
symmetries. Clearly, the general scheme can be extended to the symplectic
case with no principal difficulties.

\subsubsection{Unitary symmetry.} %
This case corresponds to considering the limit $y\to\infty$. The nonzero
contribution comes then from the second line in (\ref{f}) where one can use
effectively $4y^2e^{-2y^2(\lambda_2^2-1)}\to\delta(\lambda_2-1)$ in the
limit under discussion. This simplifies expression (\ref{Fgeneric}) further
to
$\mathcal{F}(\tilde{x}) = \int_{1}^{\infty}\!\rmd\lambda_1
\!\int_{-1}^{1}\frac{\rmd\lambda_0}{(\lambda_1-\lambda_0)^2}
\frac{\tilde{x}+\lambda_0 }{ \tilde{x}+\lambda_1}
\,e^{-\gamma(\lambda_1-\lambda_0)/2}\,.
$ 
Performing now analytical continuation (\ref{continuation}) and making use of
$\mathrm{Im\,}\frac{1}{\lambda_1-x+i0}=-\pi\delta(\lambda_1-x)$, one gets
readily $F(x)=\pi\int_{-1}^{1}\!\rmd\lambda_0(x-\lambda_0)^{-1}
e^{-\gamma(x-\lambda_0)/2}$ that yields the distribution
\cite{Fyodorov2003i}
\begin{equation}\label{Pgue}
P_0^{\mathrm{gue}}(x)=\frac{1}{2} \frac{\rmd}{\rmd x} (x^2-1) \frac{\rmd
}{\rmd x}\int_{x-1}^{x+1} \frac{\rmd t}{t} e^{-\gamma t/2}
\end{equation}
leading to equation (\ref{P(x)gue}) with $\beta=2$ obtained earlier
\cite{Beenakker2001} by a different method.

\subsubsection{Orthogonal symmetry.} %
This case amounts to investigating (\ref{f}) and (\ref{Fcross}) at $y=0$.
Fortunately, further simplifications are possible if one considers the
integrated probability distribution $W(x)\equiv-\frac{1}{2\pi}(x^2-1)
\frac{\rmd}{\rmd x}F(x) = \int_x^{\infty}\rmd x\,P_0(x)$, which is a
positive monotonically decaying function by definition. To this end, we note
that it is useful to switch to the parametrization of
\cite{Verbaarschot1985} to carry out the threefold integration. The latter
turns out to yield then a sum of decoupled terms and, after some algebra
(see Appendix C), the result can be cast \cite{Savin2005} in the following
final form:
\begin{equation}\label{Wgoe}
\fl W(x) = \frac{x+1}{4\pi}\Bigl[f_1(w)g_2(w)+f_2(w)g_1(w)
+h_1(w)j_2(w)+h_2(w)j_1(w)\Bigr]_{w=\case{x-1}{2}}
\end{equation}
with auxiliary functions defined as follows:
\begin{eqnarray}
f_1(w) &=& \int_w^{\infty}\!\rmd t
\frac{\sqrt{t|t-w|}\,e^{-\gamma{t}/2}}{(1+t)^{3/2}}
[1-e^{-\gamma}+t^{-1}]\,,
\nonumber\\
g_1(w) &=& \int_w^{\infty}\!\rmd t \frac{1}{\sqrt{t|t-w|}} \frac{e^{-\gamma
t/2}}{(1+t)^{3/2}}\,,
\nonumber\\
h_1(w) &=& \int_w^{\infty}\!\rmd t \frac{\sqrt{|t-w|}\,e^{-\gamma
t/2}}{\sqrt{t(1+t)}} [\gamma+(1-e^{-\gamma})(\gamma t-2)]\,,
\nonumber\\
j_1(w) &=& \int_w^{\infty}\!\rmd t \frac{1}{\sqrt{t|t-w|}} \frac{e^{-\gamma
t/2}}{\sqrt{1+t}}\,; \nonumber
\end{eqnarray}
their counterpart with the index 2 being given by the same expression save
for the integration region $t\in[0,w]$ instead of $[w,\infty)$. The
interpolation expression (\ref{P(x)gue}) with $\beta=1$ compared to the
above exact result shows systematic deviations in the nonperturbative regime
of moderate absorption $\gamma\sim1$ (see \cite{Savin2005}). Both formulae
essentially coincide in the limiting cases of weak or strong absorption,
when one can already use the more simple and physically transparent exact
limiting statistics (\ref{P(x)lim}).

\subsubsection{Impedance, reactance and the LDoS.}

Now we discuss statistics of the real and imaginary parts of the local Green
function one by one. Let us consider first the distribution of the imaginary
part $v$ (LDoS). Having $P_0(x)$ at our disposal, it is immediate to find
the distribution function $\mathcal{P}_{v}(v)$ of $v$ by integrating out $u$
in (\ref{P(u,v)}):
\begin{equation}\label{P(v)}
\mathcal{P}_v(v) = \frac{\sqrt{2}}{\pi v^{3/2}}\int_0^{\infty}\!\rmd q\,
P_0\left[q^2+\frac{1}{2}\Bigl(v+\frac{1}{v}\Bigr)\right]\,.
\end{equation}
This distribution is normalized to 1 and has the first moment unity at
arbitrary absorption (indeed $\langle{v}\rangle\equiv\int_{0}^{\infty}\!\rmd
v\,v{\mathcal P}(v)=1$ is automatically satisfied due to invariance of the
integrand of (\ref{P(v)}) with respect to the change $v\to\case1v$).
Explicit results for $\mathcal{P}_v(v)$ as well as for $\mathcal{P}_u(u)$
(see (\ref{P(u)}) below) can be readily obtained as $P_0(x)$ at arbitrary
$\gamma$ has been already derived above. We refer the reader to the original
publication \cite{Fyodorov2004ii}, focusing now on the physically
interesting limiting cases of weak and strong absorption, where the behavior
of the distributions is qualitatively different.

At $\gamma\ll1$, the exact limiting statistics (\ref{P(x)lim}) for $P_0(x)$
can be used to perform the integration in (\ref{P(v)}). One arrives at the
following result:
\begin{equation}\label{P(v)weak}
{\mathcal P}_{v}(v) \propto \left\{
\begin{array}{ll}
\alpha^{(1+\beta)/2}v^{-(3+\beta)/2}e^{-\alpha/4v}\,,\quad &
\!v\ll\alpha\\[1ex]
\alpha^{1/2}v^{-3/2}\,, \hfill & \alpha\ll v \ll 1/\alpha\\[1ex]
\alpha^{(1+\beta)/2}v^{-(3-\beta)/2}e^{-\alpha v/4}\,,\quad &
\!\!\!1/\alpha\ll v \,,
\end{array}\right.
\end{equation}
where constant factors of the order unity  are omitted. This result can be
physically interpreted in the single-level approximation
\cite{Efetov1993,Taniguchi1996}, when the main contribution to the LDoS
$v\approx v_n=|\psi_n|^2
\frac{\gamma}{2\pi}(\varepsilon_n^2+\frac{1}{4}\gamma^2)^{-1}$ comes from a
level closest to the given value of the energy
($\varepsilon_n=2\pi(E-E_n)/\Delta$). Fluctuations of the wave function
$\psi_n$ are mainly responsible for the exponential suppression of the
distribution on the tails, while spectral fluctuations produce the
intermediate bulk behavior with a ``super-universal'' $\frac{3}{2}$--power
law, which actually does not depend on the symmetry at all. As $\gamma$
increases, a number of levels contributing to $v\sim\sum v_n$ also grows
$\sim\gamma$ and their contributions get less correlated. The resulting
limiting form is almost a Gaussian:
\begin{equation}\label{P(v)strong}
{\mathcal P}_{v}(v) = \sqrt{\frac{\alpha}{4\pi v^{3}}}\,\exp\left[
-\frac{\alpha}{4}\Bigl(\sqrt{v}-\frac{1}{\sqrt{v}}\Bigr)^2\right]
\,,\qquad  \alpha\gg1\,.
\end{equation}
It shows a peak at $v\sim1$ of the width
$\propto1/\sqrt{\gamma}\ll1$, in agreement with
\cite{Taniguchi1996}.

\begin{figure}\label{fig2}
\includegraphics[width=\textwidth]{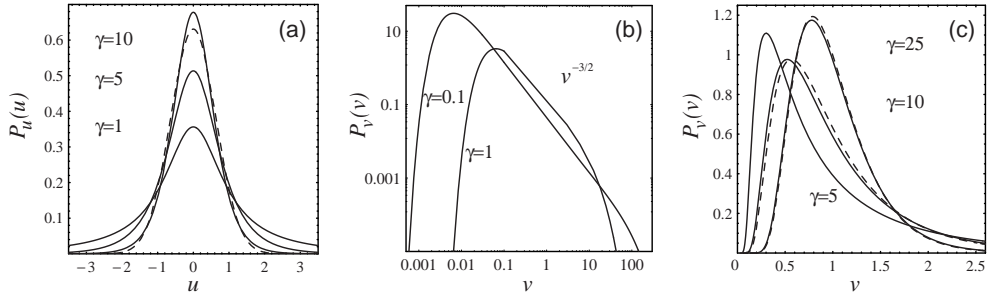}
\caption{Distribution of the real $u$  and imaginary $v$ parts of the local
Green function in the time-reversal chaotic cavity as absorption $\gamma$
grows. Dashed lines in (a) and (c) correspond to the limiting statistics
(\ref{P(u)lim}) and (\ref{P(v)strong}), respectively.}
\end{figure}
Along the same lines, we may consider the distribution $\mathcal{P}_u(u)$ of
the real part $u$ (``reactance''). One finds from (\ref{P(u,v)})
\begin{equation}\label{P(u)}
\mathcal{P}_u(u) = \frac{1}{2\pi\sqrt{u^2+1}}\int_0^{\infty}\!\rmd q\,
P_0\left[\frac{\sqrt{u^2+1}}{2}\Bigl(q+\frac{1}{q}\Bigr)\right]\,.
\end{equation}
The limiting forms of ${\mathcal P}_u(u)$ at weak and strong absorption
follow readily as
\begin{equation}\label{P(u)lim}
{\mathcal P}_{u}(u) \simeq \left\{ \begin{array}{ll}
\pi^{-1}(1+u^2)^{-1}\,, & \alpha\ll1 \\[2ex]
\sqrt{\alpha/4\pi}\,e^{-\alpha u^2/4}\,, & \alpha\gg1
\end{array} \right. \,.
\end{equation}
The Lorenzian distribution in the first line of (\ref{P(u)lim}) is the
consequence of the uniform distribution of the scattering phase
\cite{Mello1995}, as $u\approx\cot\case\theta2$ in the limit considered. As
absorption grows, one can see the crossover to a Gaussian distribution
centered at zero, which is again a result of the central limit theorem. This
type of behavior as well as a trend of $\mathcal{P}_v(v)$ to the Gaussian
distribution (\ref{P(v)strong}), see figure 2, was recently observed in
experiments on the cavity impedance \cite{Hemmady2005,Hemmady2005i}.

\subsubsection{Delay time at vanishing absorption and
eigenfunction intensity.}

The Wigner's time-delay is one of the most important and frequently used
characteristics of quantum scattering. It was recently realized
\cite{Ossipov2005} that the fundamental relation (\ref{P(u,v)}) allows one
to relate the distribution of the time-delay $q$ in single-channel
scattering (i.e. the Wigner's time-delay) from a general \emph{lossless}
disordered system to the distribution of wave function intensities in the
closed counterpart of such a system. Namely, consider the local
eigenfunction intensity $y=\mathcal{V}|\psi_n({\bf r})|^{2}$ at a spatial
point ${\bf r}$ of the closed system, with $\mathcal{V}$ denoting the volume
of the sample and the index $n$ numbering different eigenfunctions. Let
${\cal P}_y(y)$ stand for the distribution of this intensity, where the
statistics is sampled both over various realizations of the disorder in the
system and over a certain small energy range around the point $E$ in the
spectrum, with $\Delta$ being the mean level spacing in that energy range.
Define the dimensionless time delay $\tau_w=q\Delta/2\pi$ corresponding to
wave scattering at the chosen energy $E$ from a single channel attached to
the point ${\bf r}$ of the sample. Denoting ${\cal P}_w(\tau_w)$ the
corresponding distribution in the regime of perfect coupling to the sample,
Ossipov and Fyodorov managed to derive the following functional relation
between the two distributions:
\begin{equation}\label{9}
\mathcal{P}_{w} (\tau_w)=\tau_w^{-3}\,\mathcal{P}_y(\tau_w^{-1})
\end{equation}
Referring the interested reader for the details of derivation to
\cite{Ossipov2005}, we would like only to mention that at the starting point
of the derivation the time delay is expressed via the reflection coefficient $r$
in presence of absorption: $\tau_w=\lim_{\gamma\to 0}\frac{1-r}{\gamma}$,
see (\ref{R}). Then one exploits in a clever way the scaling form of the
distribution (\ref{P(u,v)}) remembering both the interpretation of the variable
$x$ in terms of $r$, see (\ref{r-x}), and the interpretation of $v$ as the LDoS,
the latter step providing a connection to eigenfunction statistics.

On one hand, in the idealized situation of zero absorption statistics of
delay times of all sorts were studied intensively in the framework of the
RMT approach, various exact analytical results being available, see
\cite{Fyodorov1997,Lehmann1995b,Gopar1996,Fyodorov1997i,%
Brouwer1997,Brouwer1999,Schomerus2000,Gopar1998,Savin2001,Sommers2001}
and references therein. Those were successfully verified in
numerical simulations of chaotic systems of quite a diverse
nature, see \cite{Glueck1999,Kottos2003,Akguc2003}. Since phase
shifts and delay times are experimentally measurable quantities,
especially in a single-channel reflection experiment
\cite{Doron1990,Mendez-Sanchez2003,Kuhl2005,Genack1999,Chabanov2001,Jian2003,Pearce2003},
the relation (\ref{9}) opens a new possibility for experimental
study of eigenfunctions.

On the other hand, one can use the existent knowledge on eigenfunction
statistics \cite{Fyodorov1995i,Mirlin2000} to provide via (\ref{9}) explicit
expressions for time-delay distributions. In this way one can e.g. recover
those for chaotic systems of all symmetry classes obtained previously by
diverse methods in various regimes of interest
\cite{Fyodorov1997,Gopar1996,Fyodorov1997i}. Of particular interest is the
predicted multifractal behaviour of the negative moments of time-delays in
the vicinity of the Anderson localisation transition \cite{Ossipov2005}:
\begin{equation}\label{mfrac}
\left\langle \tau_w^{-q} \right\rangle\propto L^{-qD_{q+1}}
\end{equation}
where $L$ stands for the system size at criticality, and $D_q$ are anomalous
(multifractal) dimensions of the eigenfunctions. Such a behaviour was indeed
discovered recently \cite{Mendez-Bermudez2005} in numerical simulations of
the disordered lattice Hamiltonians at criticality.

\subsubsection{Anderson transition as phenomenon of spontaneous
breakdown of S-matrix unitarity.}

Actually, absorption in disordered systems plays not only purely technical,
but also a conceptually important role in revealing the mechanisms behind
the Anderson localisation transition. Lets us shortly discuss a possible
qualitative behaviour of the PNR $\tau=1-r$ in a scattering system formed by
a single perfect channel attached to a d-dimensional disordered sample at
the vicinity of the point of the Anderson delocalisation transition $\mu_c$
(the mobility edge). Here we denote by $\mu$ an effective parameter which
controls the transition in the infinite sample, with states being localised
(extended) for $\mu>\mu_c$ (respectively, $\mu<\mu_c$).

For a sample of finite size $L$ the PNR is a function of three parameters:
absorption $\Gamma$, size $L$, and disorder strength $\mu$. In the
insulating phase $\mu>\mu_c$  the system is characterized by a localisation
length $\xi$ which diverges when $\mu$ approaches the critical value. A
natural scale for the absorption is played by a parameter $\delta \propto
\Gamma \xi^d$ , i.e. the ratio of the imaginary shift in energy to the mean
level spacing for sample of a typical size $L\propto \xi$ (localisation
volume).

Consider the weak absorption limit $\delta\to 0$. It is clear that the wave
incoming through the incident channel effectively explores only a part of
the sample of the order of localisation volume $\xi^d$, being exponentially
small elsewhere. Under this condition it is clear that the two limits:
$\Gamma\to 0$ and $L\to\infty$ should actually commute and can be taken in
arbitrary order. We know that in the limit $\Gamma\to 0$ the PNR behaves as
$\tau\approx \Gamma q\to 0$, see (\ref{R}). Moreover, exploiting the
relation (\ref{9}), and remembering that $\langle |\psi_n({\bf
r})|^{2k}\rangle \propto \xi^{-(k-1)d}$ it is easy to see that all negative
integer PNR moments should behave in the infinite volume limit $L\to \infty$
as $\langle \tau^{-k} \rangle \sim \Gamma^{-k} \xi^{-d k}$. With a little
bit more work one can suggest a qualitative picture for the PNR probability
distribution in the localised phase. Namely, the distribution should have a
powerlaw tail ${\cal P}(\tau)\propto \delta/\tau^2$ extending through a
parametrically large domain $\delta < \tau < 1$, and should decay very fast
towards zero for both $\tau \ll \delta$ and $\tau\to 1$. When absorption
vanishes $\delta\propto \Gamma \to 0$ such a distribution collapses to the
Dirac $\delta-$function: $\mathcal{P}(\tau)\propto \delta(\tau)$, but in a
very nonuniform way. We may conclude that in the limit of vanishing
absorption $S$-matrix unitarity is indeed recovered, and in this sense we
can associate the localised phase with the phase of unbroken symmetry.

In contrast, in the delocalized phase the incoming wave explores the whole
sample volume. It is natural to think that whatever small (but fixed) is an
absorption rate $\Gamma$, in the limit $L\to\infty$ a finite portion of the
incoming flux will be absorbed in the sample and will never come back to the
incident channel. In particular, we may expect $\lim_{\Gamma\to 0}\lim_{L\to
\infty}\tau(\Gamma,L,\mu)=\tau_{\infty}(\mu)>0$ as long as $\mu<\mu_c$. From
this point of view the Anderson transition acquires a natural interpretation
as the phenomenon of {\it spontaneous} breakdown of $S$-matrix unitarity.
Informal arguments in favour of such a behaviour are based on a picture of
the transition as described in terms of a functional order parameter
developed in detail in \cite{Mirlin1994i,Mirlin1994ii}, see also earlier
results in \cite{Efetov} and \cite{Zirnbauer1986b}. Namely, the distribution
function ${\cal P}(\tau)$ is expected to remain a non-trivial finite-width
distribution even when $\Gamma\to0$, provided the latter limit is taken {\it
after} the infinite volume limit $L\to\infty$. Clearly, more work is needed
to substantiate this claim, as well as to clarify critical behaviour of
$\tau_{\infty}(\mu)$ as long as $\mu\to\mu_c$.

Note finally, that if  the limit $\Gamma\to 0$ is taken first,
then for $\mu <\mu_c$ PNR in large but finite sample should again
scale with the system size $L$ as $\langle \tau^{-k} \rangle \sim
C(\mu) \Gamma^{-k} L^{-d k}$, cf. (\ref{mfrac}). The coefficient
$C(\mu)$ is however expected to diverge when $\mu \to \mu_c$. This
divergence should be related to the properties of eigenfunctions
via equation (\ref{9}).

\subsubsection{Arbitrary coupling to the channel.}

The general case of arbitrary transmission coefficient, $T<1$, can
be mapped \cite{Mello1985,Brouwer1995,Savin2001,Kuhl2005} to that
of perfect coupling by making use of the following relation
\begin{equation}\label{Smap}
S_{T=1}=\frac{S-\sqrt{1-T}}{1-\sqrt{1-T}S}
\end{equation}
between the corresponding scattering matrices. Equation
(\ref{Smap}) is known from the Poisson kernel theory
\cite{Mello1985}. Due to an additional interference between
incoming and directly back-scattered waves, the scattering phase
$\theta$ acquires a non-uniform distribution and acquires
statistical correlations with $x$ (or $r$). However, the joint
probability density $P(x,\theta)$ is again determined by the
function $P_0$ as follows \cite{Fyodorov2004ii} :
\begin{equation}\label{P(x,theta)}
\fl P(x,\theta) = \frac{1}{2\pi} P_0(xg-\sqrt{(x^2-1)(g^2-1)}\cos\theta) \,,
\qquad g=\frac{2}{T}-1\ge1
\end{equation}
This relation can be obtained by a straightforward evaluation in
the parametrization (\ref{Smap}) of the corresponding Jacobian.
The integration over $x$ immediately yields the scattering phase
distribution. In the particular case of vanishing absorption
$r\to1$ and $P_0(x)\to x^{-2}\delta(1/x)$. This readily gives
$P(x,\theta)\to\rho(\theta)P_0(x)$, with the phase density
$2\pi\rho(\theta)=(g-\sqrt{g^2-1}\cos\theta)^{-1}$ found earlier
\cite{Savin2001} (see \cite{Akguc2003} for the corresponding
numerical study). As another example we consider the reflection
coefficient distribution in terms of $P_0(x)$ is given at
arbitrary $T$ by \cite{Fyodorov2003i}:
\begin{equation}\label{P(r)}
P_{r}(r) = \frac{1}{\pi(1-r)^2}\int_0^{2\pi}\!\!\rmd\theta P_0
\Bigl[\frac{2(g-\sqrt{g^2-1}\sqrt{r}\cos\theta)}{1-r}-g\Bigr] \,.
\end{equation}
Distributions of $r$ and $\theta$ in microwave cavities were recently
studied for different realizations of $\gamma$ and $T$ in \cite{Kuhl2005},
the excellent agreement with the theory being found.

\subsection{Beyond single channel}
\subsubsection{Reflection coefficients and PNRs.}

The starting point of our analysis in this section is the following
convenient representation \cite{Fyodorov2003i} for the diagonal elements
$S^{cc}(E)$ of the scattering matrix, cf.~(\ref{S}):
\begin{equation}\label{Scc}
S^{cc}(E) = \frac{1-\rmi K_c(E)}{1+\rmi K_c(E)}\,, \qquad
K_{c}(E)=\case12V^{c\dag}(E-\Heff^c)^{-1}V^c\,.
\end{equation}
where $\Heff^c=H-\case\rmi2\sum_{b\ne c}V^bV^{b\dag}$ is now the
$c-$dependent non-Hermitian operator. In view of
$\Heff=H-\frac{i}{2}VV^{\dagger}\equiv\Heff^c-\frac{i}{2}V^cV^{c\dagger}$
one can treat $V^cV^{c\dagger}$ as a rank one perturbation with
respect to $\Heff^c$. In this case the following general
relationship (Dyson's equation) is valid for the corresponding
resolvents \cite{Sokolov1989}:
\begin{equation}\label{Dyson}
\fl \frac{1}{E-\Heff}=\frac{1}{E-\Heff^c}-\frac{\rmi}{2}\frac{1}{E-\Heff^c}
V^c\frac{1}{1+\rmi K_c(E)}V^{c\dagger}\frac{1}{E-\Heff^c}
\end{equation}
which can be proved by expanding the l.h.s. in a power series with
respect to $(E-\Heff^c)^{-1}V^cV^{c\dagger}$ and summing then up
the resulting geometric series in $K_c$. Substituting this
equation in $S^{cc}=1-\rmi V^{c\dagger}(E-\Heff)^{-1}V^c$ one gets
$S^{cc}=1-2\rmi K_c-2K_c(1+\rmi K_c)^{-1}K_c$ which is equivalent
to (\ref{Scc}). It is also worth noting that a representation like
(\ref{Scc}) is valid for an arbitrary $m{\times}m$ submatrix
standing on the main diagonal of $S$ with obvious replacement
$V^{c}$ by the $N{\times}m$ matrix
$V^{(m)}=(V^{c_1},\ldots,V^{c_m})$, and corresponding changes for
$K_c$ and $\Heff^c$. In the particular case $m=M$, i.e. the full
$S$-matrix, one recovers (\ref{S}) from (\ref{Heff}) and
(\ref{Dyson}).

The relation (\ref{Scc}) reduces the problem of evaluating the statistics of
$S_{cc}$, and hence the reflection coefficient / PNR in a given channel $c$
to calculating the joint probability density $\mathcal{P}(u_c,v_c)$ of
$u_c=\mathrm{Re\,}K_c$ and $v_c=-\mathrm{Im\,}K_c$, with $K_c$ standing for
the particular diagonal entry of the resolvent of the effective {\emph
non-Hermitian} Hamiltonian $\Heff^c$ from (\ref{Scc}). Moreover,  a uniform
absorption within the sample can again be taken into account by a purely
imaginary shift of the scattering energy $E\to E+\case\rmi2\Gamma\equiv
E_{\gamma}$, all further steps being fully in parallel to those of section
4.1.2. The most pleasant feature of this approach is that it is very
straightforward to include open channels  in the derivation of
$\mathcal{P}(u_c,v_c)$ within the supersymmetry method. All important
properties of this distribution, in particular, the relations
(\ref{P(u,v)}), (\ref{Fcross}) and (\ref{Pgeneric}) retain its validity,
with $x$ being naturally replaced by $x_c=\case1{2v_c}(u_c^2+v_c^2+1)$. The
corresponding function $F(x_c)$ then follows from (\ref{Fcross}) by
multiplying there the integrand with the ``channel factor'' (which
originates from the imaginary part of $\Heff^c$)
\begin{equation}\label{Fchan}
\fl \mathcal{F}_M^c(\{\lambda\})=\prod_{b\neq
c}\left[\frac{(g_b+\lambda_0)^2 }{
(g_b+\lambda_1\lambda_2)-(\lambda_1^2-1)(\lambda_2^2-1)}\right]^{1/2},
\qquad g_b=\frac{2}{T_b}-1
\end{equation}
which accounts for \emph{arbitrary} coupling to all other channels save for
the given perfectly open one $c$.  Arbitrary coupling for the remaining
channel $c$ can be considered by means of (\ref{P(x,theta)}), providing us
finally with the general distribution of the reflection amplitude and phase.
In this way one can also obtain explicit distributions for many interesting
situations \cite{Fyodorov2003i}, including the cases when the effects of
Anderson localisation start to play an important role. In particular, in the
case of broken TRS one can calculate the PNR distribution for a single
channel attached to an edge of a piece of quasi-one-dimensional disordered
medium of a given length $L$, with the opposite edge being either closed, or
in contact with a perfectly conducting multichannel lead.

\subsubsection{Reflection eigenvalues and thermal emission.}
The exact result for the distribution function $P(r)$ of
reflection eigenvalues valid for any number of arbitrary open
channels and arbitrary absorption is known only for the $\beta=2$
case, being recently calculated by two of us \cite{Savin2003i},
generalizing earlier perturbative results \cite{Beenakker2001}
(known for all $\beta$). That uses the method developed for
studying the proper time-delay distribution \cite{Sommers2001} in
the ideal lossless system. The later distribution considered at
finite absorption by means of (\ref{R}) has a sharp cutoff at
$q=\Gamma^{-1}$  due to a finite value of $P(r)$ at $r=0$. This
fact makes the interpretation of $q$ as delay times at strong
absorption questionable \cite{Savin2003i},  since intuitively one
expects a generic exponential suppression at large values of delay
times $q\gg\Gamma^{-1}$. Indeed, for the time $\delta t$ a
wave-packet oscillating in the cavity with a frequency
$\Delta/2\pi$ on average experiences $(\Delta/2\pi)\delta t$
collisions with the walls, yielding the probability
$T_{\phi}(\Delta/2\pi)\delta t$ to be absorbed into one of
$M_{\phi}$ parasitic channels ($T_\phi\ll1$ being the transmission
coefficient). The total reflection is then estimated as
$R\simeq(1-T_{\phi}(\Delta/2\pi)\delta t)^{M_{\phi}}$, giving
$e^{-\Gamma\delta t}$ in the absorption limit of the fixed
absorption width $\Gamma=M_{\phi}T_{\phi}\Delta/2\pi$ as
$M_{\phi}\to\infty$ and $T_{\phi}\to0$. It is natural, therefore,
to define alternatively the positive definite matrix $Q_R$ of
\emph{reflection} time-delays as follows \cite{Savin2003i}:
\begin{equation}\label{QR}
Q_R \equiv-\Gamma^{-1}\ln R_{\gamma}= -\Gamma^{-1} \ln(1-\Gamma
Q_{\gamma})\,.
\end{equation}
One finds easily the connection $\mathcal{P}_R(q_r) = e^{-\Gamma q_r}
\mathcal{P}[\frac{1}{\Gamma}(1-e^{-\Gamma q_r})]$ between the corresponding
distributions $\mathcal{P}_R(q_r)$ and $\mathcal{P}(q)$ of proper delay
times (eigenvalues of $Q_R$ and $Q_{\gamma}$, respectively). Both $Q_R$ and
$Q_{\gamma}$ reduce to the same Wigner-Smith matrix (\ref{Q}) in the limit
of vanishing absorption. The difference between them becomes noticeable at
finite $\Gamma$. Still both distributions coincide up to the time
$\ll\Gamma^{-1}$. They start to differ at larger times, when
$\mathcal{P}(q)$ has the cutoff whereas
$\mathcal{P}_R(q\gg\Gamma^{-1})\propto e^{-\Gamma q}$ is exponentially
suppressed.

The exact result for the $\beta=1$ case at arbitrary $M$ is still
outstanding. In the limit of the large number of equivalent channels,
$M\to\infty$, an exact result can be found at arbitrary absorption and
coupling \cite{Savin2004}, and the perfect coupling case $T=1$ has been
known for some time \cite{Beenakker1998,Beenakker1999}. In contrast to the
few-channel case, when any value $0\le r \le 1$ is permitted, the
distribution density in the present case is non-vanishing only in a range
$0\le r_{\mathrm{min}}\le r\le r_{\mathrm{max}}<1$, and is the same for all
$\beta$. Referring the reader to \cite{Savin2004} for explicit results, we
mention their application for thermal emission from random media.
Registration of $n$ photons in the frequency window $\delta\Omega$ during
the large time $t\gg1/\delta\Omega$ yields the negative-bimodal distribution
of photocounts with $\nu=Mt\delta\Omega/2\pi$ degrees of freedom; see
\cite{Mandel}. In his seminal paper Beenakker \cite{Beenakker1998} has shown
that the quantum optical problem of the photon statistics can be reduced to
a computation of the $S$ matrix of the classical wave equation. In
particular, chaotic radiation may be characterized by the effective number
$\nu_{\mathrm{eff}}$ degrees of freedom as follows \cite{Beenakker1998}:
$\nu_{\mathrm{eff}}/\nu=(1-\langle{r}\rangle)^2/\langle(1-r)^2\rangle\leq1$,
with $\nu_{\mathrm{eff}}=\nu$ for blackbody radiation. This ratio is given
by $\nu_{\mathrm{eff}}/\nu=(\gamma_s+T)^2/[\gamma_s^2+2(\gamma_s+T)]$ at
arbitrary transmission $T$  and absorption $\gamma_s\equiv2\pi\Gamma/\Delta
M$, implying thus that saturation to the blackbody limit slows down,
$\frac{\nu_{\mathrm{eff}}}{\nu}\approx1-\case2\gamma_s(1-T)$ at
$\gamma_s\gg1$, for transmission $T<1$ \cite{Savin2004}. Finally, due to a
duality relation \cite{Beenakker1998,Paasschens1996} of a linear amplifying
system to a dual absorbing one (related to it by inverting the sign of
$\Gamma$) there exists a further link of the analysis presented here to the
rapidly developing field of random lasers
\cite{Beenakker1998,Beenakker1999,Paasschens1996,Hackenbroich2001,Cao2003}.

\subsubsection{Off-diagonal entries of the Green function.}

As we have seen, the statistics of real and imaginary parts of diagonal
elements of the Green function can be very efficiently studied in the
framework of the supersymmetric approach and may have various physical
interpretations. The off-diagonal entries $G _{ij}(E_\gamma)=\langle
i|(E_\gamma-\Heff)^{-1}|j\rangle$ are of considerable importance as well. It
can be easily understood that $W=|G_{ij}|^2$ is essentially the wave power
transmitted from a source at site $i$ inside a random medium to a receiver
at site $j$. Statistics of such an object is much more difficult to study in
general. Presently the most studied case \cite{Rozhkov2003} is $\beta=2$
symmetry class under an additional assumption that both receiver and source
are very weakly coupled to the medium as to ensure those couplings do not
contribute essentially to the resonance broadening. This means the
broadening is induced purely due to losses elsewhere in the medium.
Technically the latter requirement amounts to vanishing coupling amplitudes
$V^c_i$ and $V^c_j$ for all damping channels $c=1\ldots M$. Assuming further
the RMT statistics for $\mathrm{Re\,}\Heff$, it is easy to see that the
damping matrix $\mathrm{Im\,}\Heff$  can be chosen diagonal and simply such
that both entries $(VV^{\dagger})_{ii}$ and $(VV^{\dagger})_{jj}$ are
vanishing. For such a model the distribution of the scaled transmitted power
$w=(\case\Delta\pi)^2W$ can be found explicitly \cite{Rozhkov2003} as:
\begin{equation}\label{power}
\fl {\cal P}(w)=\left(\frac{\rmd}{\rmd w} + w\frac{\rmd^2}{\rmd w^2}
\right)\left[\frac{e^{-\gamma\sqrt{1+w}}}{2\sqrt{1+w}}
\int_{-1}^1d\lambda\frac{\sqrt{1+w}+\lambda}
{\sqrt{1+w}-\lambda}e^{\gamma\lambda}
\prod_{c=1}^M\frac{g_c+\lambda}{g_c+\sqrt{1+w}}\right]\,.
\end{equation}
In fact, the remaining integration can be performed for two
interesting cases: (i) all equivalent dissipation channels $g_c=g$
in the absence of uniform losses $\gamma=0$ and (ii) no internal
channels of dissipation $g_c\to \infty$ in the presence of uniform
absorption $\gamma>0$. Here we present the formula only for the
latter case \cite{Rozhkov2003}:
\begin{equation}\label{power1}
\fl \mathcal{P}(w)=\frac{\gamma^4e^{-\gamma_w}\sinh{\gamma}}{4\gamma_w^5}
\left[ \gamma_w^2(w+2)
-(w-2-2(\gamma_w^2/\gamma)\coth{\gamma})(1+\gamma_w)\right]
\end{equation}
with the shorthand $\gamma_w\equiv\gamma\sqrt{1+w}$. The distribution of the
transmitted power for the case of absorptive media with preserved
time-reversal invariance is not yet known. First two moments of that
quantity were calculated recently in \cite{Rozhkov2004}.

Let us finally mention that a closely related question of statistics of
intensity of waves emitted from a permanently radiating source embedded in a
random medium was the subject of many studies in recent years. Some useful
references can be found in Section 7 of \cite{Mirlin2000}, see also the
relevant review \cite{Cao2003} on random lasing.

\section{Conclusions and open problems}

For wave scattering in open chaotic and/or disordered systems with uniform
losses, we have discussed various statistics on the level of both
correlation and distribution functions. The overall exponential decay due to
uniform absorption is the generic feature of any correlation
function reduced to a two-point spectral (resolvent) correlation function,
that follows simply from analytic properties of the latter in the complex
energy plane. For fully chaotic systems with or without TRS, we have
calculated exactly energy correlation functions of complex impedance and
$S$-matrix elements at arbitrary absorption and coupling. The corresponding
enhancement factors have been also discussed in detail. The result for
$S$-matrix correlations in the case of broken TRS completes the well known
one \cite{Verbaarschot1985} of preserved TRS.

To study distribution functions, we have described the novel approach to the
problem by deriving a kind of a fluctuation dissipation relation
(\ref{Phat}) of quite a general nature, which relates distributions at
finite absorption to arbitrary order correlations at zero absorption via a
nontrivial analytic continuation procedure. Correlations can be efficiently
studied in the framework of the supersymmetric nonlinear $\sigma$-model. In
the zero dimensional case, a number of explicit results is provided for
quantities characterizing open chaotic systems both from ``inside'' (LDoS,
the complex impedance and the (local) Green's function) and from ``outside''
(PNRs, reflection coefficients and time delays). This $\sigma$-model mapping
provides an attractive possibility to include into consideration open
disordered absorptive systems beyond the usual RMT treatment.

Finally, let us briefly discuss an (incomplete) list of problems deserving,
from our personal point of view, further investigations both for scattering
systems with and without absorption. Within the domain of RMT applicability,
the most outstanding problems are of course those requiring evaluation of
four-point correlation functions of the resolvents, e.g., partial scattering
cross-sections. These are still not known even for the simplest case of
broken TRS. Another challenge is to find a probability distribution for the
multichannel $S$-matrix at finite absorption, thus nontrivially generalizing
the Poisson kernel \cite{Mello1995}. More work is required to understand an
interplay between the statistics of complex resonances and the corresponding
bi-orthogonal eigenfunctions \cite{Sokolov1989,Fyodorov2003}, the question
being of particular relevance for lasing from random and/or chaotic media
\cite{Schomerus2000a,Hackenbroich2002}.
Spatial characteristics of internal parts $b^c=(E-\Heff)^{-1}V^c$ of the
scattering wave function are very interesting by their own and are closely
related to fluctuations in transmitted power discussed above in 4.2.3.

Apart from that, last decades it was realized \cite{Altland1997} that it may
be required to consider other symmetry classes (beyond the three of Dyson)
which are relevant, e.g., for systems involving superconducting elements.
The corresponding scattering theory is reviewed in \cite{Beenakker2005}. To
reconsider many problems discussed in the present review taking those (and
other related) symmetries into account should be both interesting and
informative (see \cite{Fyodorov2004iii} as a recent example).

At last but not least: understanding scattering in disordered systems beyond
the universal RMT regime, taking into account, in particular, the effects of
Anderson localisation, is a rather promising area calling for more
systematic research.

\ack The work was supported by EPSRC grant EP/C515056/1 ``Random Matrices
and Polynomials'' (Y.V.F.) and SFB/TR 12 der DFG (D.V.S. and H.-J.S.).

\appendix
\section{Statistics of the local Green function in the GSE}
\setcounter{section}{1}

We start with expressing the dimensionless LDoS (the imaginary part of the
local Green's function $ G(E+\case\rmi2\Gamma; {\bf r}; {\bf r})$ in units
of $\Delta$) in terms of the eigenfunctions and eigenvalues of the
Hamiltonian $H$ with underlying simplectic symmetry as:
\begin{equation}
\fl v({\bf r})=-\frac{\Delta}{\pi}\mbox{Im}\,G(E+\case\rmi2\Gamma;{\bf r},
{\bf r})=\frac{\Delta}{\pi}
 \sum_{n=1}^N \left[|\psi_n({\bf r})|^2+|\phi_n({\bf r})|^2\right]
\frac{\Gamma}{(E-E_n)^2+\case14\Gamma^2}\,.
\end{equation}
Here $\psi_n({\bf r})$, $\phi_n({\bf r})$ stand for the local amplitudes of
the two eigenfunctions corresponding to the (double-degenerate) eigenvalue
$E_n$. It is convenient to consider scaled eigenvalues $\epsilon_n=\pi
E_n/\Delta$, defining absorption $\eta=\pi\Gamma/\Delta\equiv\case12\gamma$.
Replacing $H$ with $2N\times 2N$ random GSE matrix, we follow the idea first
suggested in \cite{Beenakker1994} and in the first step exploit the
well-known fact that in the limit $N\gg 1$ eigenvector components
$\psi_n({\bf r}), \phi_n({\bf r})$ for different values of $n$ and ${\bf r}$
can be treated as independent, identically distributed complex Gaussian
variables. The corresponding joint probability density can be written
symbolically as $ {\cal P}_n(\psi;\phi){\cal D}
(\psi;\phi)=e^{-N(\psi^{\dagger}\psi+\phi^{\dagger}\phi)}{\cal D}
(\psi;\phi)$ where we introduced a shorthand notation
$\psi^{\dagger}\psi=\sum_{n=1}^N |\psi_n({\bf r})|^2$ (and similarly for
$\phi^{\dagger}\phi$), with the measure ${\cal D}(\psi;\phi)$ being
understood as the (normalized) product of differentials of independent
variables.

The Laplace transform $F(s)=\int_0^{\infty} e^{-\rho
s}\mathcal{P}_{\rho}(\rho)\rmd\rho$ of the probability distribution function
${\cal P}_{\rho}$ for the normalized variable $\rho=N\pi v({\bf r})$ can be
then written as:
\begin{equation}
\fl F(s)=\left\langle \int {\cal P}_n(\psi;\phi){\cal D}
((\psi;\phi))\prod_{n=1}^{N}\exp{-\frac{sN\eta \left(|\psi_n({\bf
r})|^2+|\phi_n({\bf r})|^2\right)}{(\epsilon-\epsilon_n)^2+\eta^2}
}\right\rangle_{\{\epsilon_n\}}
\end{equation}
where the angular brackets stand for the averaging over the joint
probability density of all eigenvalues $\epsilon_n$ and the limit
$N\to\infty$ is to finally taken. After performing the Gaussian integrals,
we therefore arrive to the following representation:
\begin{equation}\label{f2}
F(s)=\lim_{N\to \infty}\left\langle
\prod_{n=1}^{N}\frac{\left[(\epsilon-\epsilon_n)^2 +
\eta^2\right]^2}{\left[(\epsilon-\epsilon_n)^2+\eta^2+2\eta s\right]^2}
\right\rangle_{\{\epsilon_n\}}
\end{equation}
Introducing the product
$\pi_{N}(\lambda)=\prod_{n=1}^{N}\left(\lambda-\epsilon_n\right)$ we see
that the characteristic polynomial of the GSE matrix $\hat{H}$ is simply
$\pi^2_{N}(\lambda)$. The formula (\ref{f2}) can be conveniently rewritten
in terms of this polynomial as:
\begin{equation}\label{f3}
\fl F(s)=\lim_{N\to
\infty}\left\langle\frac{\pi_N^{\beta/2}(\epsilon+\rmi\eta)
\pi_N^{\beta/2}(\epsilon-\rmi\eta)}
{\pi_N^{\beta/2}(\epsilon+\rmi\mu)\pi_N^{\beta/2}(\epsilon-\rmi\mu)}
\right\rangle_{\{\epsilon_n\}},\quad \mu=\sqrt{\eta^2+2\eta s}
\end{equation}
with $\beta=4$ for GSE. In fact, in such a form the formula retains its
validity for GUE $\beta=2$ and GOE $\beta=1$ cases as well.

The problem of evaluating ensemble averages of products and ratios of
characteristic polynomials of random matrices attracted a lot of research
interest recently, both in physical and mathematical community. For
$\beta=2$, the general problem was solved in
\cite{Fyodorov2003ii,Strahov2003}, and the most complete set of formulae
available presently for $\beta=1,4$ can be found in the recent paper by
Borodin and Strahov \cite{Borodin2005}. When addressing the most interesting
case $\beta=1$ they were, unfortunately, able to consider only integer
powers of characteristic polynomials, whereas (\ref{f3}) obviously requires
knowledge of half-integer powers. In this sense the corresponding random
matrix problem for $\beta=1$ case is still outstanding. Borodin and Strahov
derived the following explicit expression for the $\beta=4$ average
featuring in (\ref{f3}):
\begin{equation}\label{BS}
\lim_{N\to
\infty}\left\langle\frac{\pi_N^2(\alpha_1)\pi_N^{2}(\alpha_2)}
{\pi_N^2(\beta_1)\pi_N^2(\beta_2)}
\right\rangle_{GSE}=\frac{\prod_{i,j=1}^2(\alpha_i-\beta_j)}
{(\alpha_1-\alpha_2)(\beta_1-\beta_2)}\,\,\mbox{Pf}\,\hat{C}
\end{equation}
assuming that $\mbox{Im}{\beta_1}>0,\,\mbox{Im}{\beta_2}<0$. Here, the
matrix $\hat{C}$ has the following structure
$\scriptsize\hat{C}=\left(\begin{array}{cc}\hat{A}&\hat{B}
\\-\hat{B}^T&\hat{D}
\end{array}\right)$, with
$\scriptsize\hat{A}=\left(\begin{array}{cc} 0&a\\-a&0
\end{array}\right)$,
$\scriptsize\hat{D}=\left(\begin{array}{cc}0&d\\-d&0\end{array}\right)$,
$\scriptsize\hat{B}=\left(\begin{array}{cc}b_{11}&b_{12}\\b_{21}&b_{22}\end{array}\right)$
and Pf stands for the corresponding Pfaffian,
$\mbox{Pf}\,\hat{C}=\sqrt{\mbox{det}\,\hat{C}}$. The entries of the matrix
$\hat{C}$ are given explicitly by:
\begin{equation}\label{bs1}\fl
a=\frac{1}{\pi}\int_{0}^{1}\frac{\rmd t}{t}\sin{(\alpha_1-\alpha_2)t},
\qquad d=2\pi \rmi\frac{\partial}{\partial \beta_1}\left(
\frac{e^{\rmi(\beta_1-\beta_2)}}{\beta_1-\beta_2}\right)
\end{equation}
\begin{equation}\label{bs2}\fl
b_{11}= -\frac{e^{\rmi(\beta_1-\alpha_1)}}{\beta_1-\alpha_1}, \quad b_{12}=
\frac{e^{\rmi(\alpha_1-\beta_2)}}{\alpha_1-\beta_2},\quad b_{21}=
-\frac{e^{\rmi(\beta_1-\alpha_2)}}{\beta_1-\alpha_2}, \quad b_{22}=
\frac{e^{\rmi(\alpha_2-\beta_2)}}{\alpha_2-\beta_2}
\end{equation}
Restricting our consideration in (\ref{f3}) for simplicity to
$\epsilon=E=0$ we have therefore
$\alpha_1=\rmi\eta,\alpha_2=-\rmi\eta,\beta_1=\rmi\mu,\beta_2=-\rmi\mu$.
Substituting this to (\ref{bs1})--(\ref{bs2}) and then to
(\ref{BS}), we find after straightforward computations the Laplace
transform of the LDoS probability density:
\begin{equation}\label{f4}
F_{GSE}(s)=F_{GUE}(s)+\delta F(s), \end{equation}
\begin{equation}\label{f5}
F_{GUE}(s)=\frac{1}{4\eta\mu}
\left(e^{-2(\eta-\mu)}(\eta+\mu)^2-e^{-2(\eta+\mu)}(\eta-\mu)^2\right)
\end{equation}
\begin{equation}\label{f6}
\delta F(s)=\frac{1}{4\eta\mu^2}(\eta^2-\mu^2)^2
e^{-2\eta}\left(1+\frac{1}{2\mu}\right) \int_{0}^{1}\frac{\rmd t}{t}
\sinh{(2\eta t)},
\end{equation}
We note that (i) the only $s-$dependence in the above expression
comes from $\mu=\sqrt{\eta^2+2\eta s}$ and (ii) the formula
(\ref{f5}) concides with the Laplace transform of the LDoS
distributuion for $\beta=2$ case, see e.g.
\cite{Efetov1993,Beenakker1994}.

The remaining job is to find the part $\delta F(\rho)$ of the LDoS
probability density corresponding to inversion of the Laplace transform in
(\ref{f6}), which gives
\begin{equation}\label{f7}
\delta
F(\rho)=\frac{\rmd^2}{\rmd\rho^2}\left[\rho^{1/2}e^{-\eta(\rho/2+2/\rho)}\right]
\times \frac{1}{4}\sqrt{\frac{2}{\pi \eta}}\int_{0}^{1}\frac{\rmd t}{t}
\sinh{(2\eta t)}
\end{equation}
Knowledge of this function allows immediate restoration of the corresponding
probability density (\ref{P(x)gse}) for the main scaling variable $x$ due to
relations discussed in the main body of the present paper (recall that
$\eta=\gamma/2$).

\section{Disordered systems of arbitrary dimension: \\ nonlinear
$\sigma$-model derivation of expressions (\ref{Cgeneric})}

Let $\mathbb{H}$ be a (self-adjoint) Hamiltonian describing a one-particle
quantum motion in a d-dimensional static disordered potential. Among
microscopic model Hamiltonians ensuring validity of the non-linear
sigma-model description of such a system the simplest choice seems to be the
Wegner's $N-$orbital model \cite{Wegner1979,Schaefer1980}, or its variant
due to Pruisken and Sch\"{a}fer \cite{Pruisken1982}. Physically the models
are equivalent to the so-called ``granulated metal'' model \cite{Efetov}.
One can visualize it by considering a lattice of $L^d$ sites ($d$ standing
for the spatial dimension of the sample) , each site being occupied with a
metallic ``granula''. The motion of a quantum particle inside each granula
is assumed to be fully "ergodic", and as such the Hamiltonian of the
individual granula should be adequately modelled by a random $N\times N$
matrix $\hat{H}$ of appropriate symmetry, provided we consider the limit
$N\to \infty$. The quantum particle is also allowed to tunnel between the
neighbouring granulae, the process ensuring a possibility of nontrivial
diffusive motion along the lattice. Thus, the Hamiltonian $\mathbb{H}$ of
the system as a whole has a form of large matrix of the size $N\times L^d$,
consisting of coupled matrix blocks $\hat{H}_k,\,\, k=1,2,\ldots ,L^d$ of
the size $N\times N$. For example, for the simplest quasi-one dimensional
sample $d=1$ such a matrix will be ``block-three-diagonal''.

Being interested in scattering, we should provide a way to incorporate
description of external leads (or waveguides) attached to the sample. In the
framework of the present model attaching an $M_k$-channel lead to the block
at site $k$ is done  by replacing the corresponding ``intragranula'' matrix
block $\hat{H}_k,$ with its non-selfadjoint counterpart
$\hat{H}_k-\case\rmi2\hat{\Gamma}_k$, where $N\times N$ matrix
$\hat{\Gamma}_k=2\,\mbox{diag}(\kappa_1^{(k)},
\ldots,\kappa_{M_k}^{(k)},0,\ldots,0)\ge 0$, and $M_k$ staying finite when
$N\to \infty$. Assuming the absorption width $\Gamma$ to be uniform and
identical in all the granulae, we are interested in deriving the joint
probability density ${\cal P}(u,v)$ of real $u$ and imaginary $v$ parts of
the local Green's function in a state $|j^{(l)}\rangle$ belonging to granule
at the lattice site $l$:
\begin{equation}\label{G}
G(j^{(l)},j^{(l)}, E_\gamma)=\langle
j^{(l)}|(E_{\gamma}-\mathbb{H}+\case\rmi2\hat{\Gamma})^{-1}|j^{(l)}\rangle\,.
\end{equation}

Following the general strategy (see Section 4.1.2) we recover
$\mathcal{P}(u,v)$ from the correlation function $C(z',z'')$, see (\ref{C}).
The calculation of the corresponding correlation function is a
straightforward extension of the ``zero-dimensional'' procedure employed in
\cite{Fyodorov1997} for $\beta=2$ and then in \cite{Fyodorov1997i} for
$\beta=1$ (as well as for the whole crossover) to the present d-dimensional
situation and will not be repeated here. The emerging supersymmetric
nonlinear $\sigma$-model on the lattice is described in terms of the
supermatrices $Q_k,\, k=1,\ldots L^d$ parametrized as
$Q_k=-\rmi\hat{T}_k\Lambda\hat{T}_k^{-1}$ and interacting according to the
``action'' (see e.g. \cite{Mirlin1994i})
\begin{equation}\label{action}
S\{Q\}=-\frac{t\beta}{4}\sum_{<k,j>}^{L^d} \mbox{Str}\,\, Q_k
Q_j+\frac{\beta}{4}\frac{\Gamma}{\Delta} \sum_{k}^{L^d} \mbox{Str}\,\,
Q_k\Lambda
\end{equation}
where the first sum goes over pairs $<k,j>$ of nearest neighbours on the
underlying lattice, $\Delta$ stands for the mean level spacing and $t$
stands for the effective inter-granule coupling constant, which is the main
control parameter of the emerging theory. We start with outlining the
procedure for $\beta=2$ symmetry when $Q$-matrices involved in the
calculation have the smallest size $4{\times}4$. Assuming for simplicity
that the granule $l$ does not have a channel attached to it directly, the
correlation function $C(z',z'')$, see (\ref{C_omega}) and (\ref{C}), of
resolvents of the (scaled with $\case{N\Delta}\pi$) Green's functions
(\ref{G}) is expressed in terms of the $Q$-integrals as follows
\begin{eqnarray}\label{l1}
\fl C_{\beta=2}(z',z'')=\prod^{L^d}_{k=1}\int
\rmd\mu(Q_k)\,\mathcal{F}^{(k)}_{M_k}(Q_k)\, e^{-S\{Q\}}
{\textstyle\frac{\partial^2}{\partial J_1\partial J_2}}
\mbox{Sdet}^{-1}(U_J^{-1}+Q_l)\big|_{J_1=J_2=0}
\end{eqnarray}
where the supermatrix $U$ depends on the variables $z',z''$ and sources
$J_1,J_2$ via
\begin{equation}\label{U}
\fl U = \case12(z'-iz'')(1+\Lambda)+\case12(z'+\rmi z'')(1-\Lambda)+
\mbox{diag}(0,J_1,0,J_2)\,.
\end{equation}
The ``channel factor'' $\mathcal{F}^{(k)}_{M_k}(Q_k)\equiv\prod_{c=1}^{M_k}
\mbox{Sdet}^{-1}(1+\rmi\kappa_c^{(k)}Q_k\Lambda)$ originates from coupling
to continuum, where $\kappa^{(k)}=0$ if no external channels is attached to
a given granule. The factors $\mathcal{F}_{M}$ appearing explicitly in
Section 3.2 as well as in (\ref{Fchan}) are just the zero-dimensional
analogue of $\mathcal{F}^{(k)}_{M_k}$. Following
\cite{Mirlin1994i,Mirlin1994ii}, we find it convenient to introduce the
function
\begin{equation}\label{l2}
Y^{(l)}(Q_l)=\prod^{L^d}_{k\ne l}\int
\rmd\mu(Q_k)\,\mathcal{F}^{(k)}_{M_k}(Q_k)\,e^{-S\{Q\}}\,.
\end{equation}
Employing the standard Efetov-type parametrization of the matrices $Q$, one
finds that actually $Y^{(l)}(Q)$ must be a function of only two commuting
variables: $1\le \lambda_1\le \infty$ and $-1\le \lambda_0\le 1$. Then
integrating out all the remaining degrees of freedom in quite a standard way
(see \cite{Fyodorov1997} for more detail), we arrive at the representation
(\ref{Cgeneric}) where
\begin{equation}\label{l4}
\mathcal{F}_{\beta=2}(z',z'')=
\int_1^{\infty}\rmd\lambda_1\int_{-1}^{1}\frac{\rmd\lambda_0}
{(\lambda_1-\lambda_0)^2} \frac{\tilde{x}+\lambda_0}
{\tilde{x}+\lambda_1}\,Y^{(l)}(\lambda_1,\lambda_0)
\end{equation}
the main scaling variable of our theory
$\tilde{x}\equiv\case1{2z''}(z'^2+z''^2+1)$ being introduced. This
immediately verifies the structure of the important formula
(\ref{Cgeneric}) for a general d-dimensional $\beta=2$ system with
attached scattering channels. Of course, ability to work out
explicit expressions for the probability density ${\cal P}(u,v)$
crucially depends on the availability of the function
$Y^{(l)}(\lambda_1,\lambda)$. This function has a very simple form
in "zero dimension", physically equivalent to a system consisting
of a single granula:
\begin{equation}
Y^{(l)}(\lambda_1,\lambda_0)=e^{-\gamma(\lambda_1-\lambda_0)/2}\prod_{c=1}^{M}
\frac{g_c+\lambda_0}{g_c+\lambda_1}
\end{equation}
where we have assumed that the total number of external channels
attached to the system is $M$. Each channel is characterised by
its own effective coupling constant
$g_c=\frac{1}{2}(\kappa_c+\kappa_c^{-1})\ge1$, the transmission
coefficient being $T_c=\case2{g_c+1}$. Weak-localisation
corrections to zero dimensional results can be in principle found
systematically following the procedures of \cite{Fyodorov1995i},
and non-perturbative results are mainly available in quasi-1D
systems, see e.g. \cite{Fyodorov1994}, in the limit of weak
absorption, $\gamma\ll1$.

Treating systems of $\beta=1$ symmetry class follows exactly the same steps
as outlined before, although in this case the supermatrices $Q$ are of the
size $8{\times}8$ with the corresponding doubling the dimension of
(\ref{U}). Following \cite{Fyodorov1997i,Altland1992}, we consider the whole
crossover between $\beta=1$ and $\beta=2$ symmetry classes. Due to the TRS
breaking perturbation controlled by the parameter $y$ ($\hat{H}_k$ is no
longer symmetric but still Hermitian), the action (\ref{action}) (with
$\beta=1$) acquires additionally the symmetry breaking term
$S_y\{Q\}=\case14y^2\sum_{k}\mbox{Str}\,(\tau_3Q_k)^2$, with $\tau_3$ being
the Pauli matrix. As usual, explicit integrations are much more cumbersome
and require more work, some very useful and helpful relations can be found
in \cite{Altland1992}. Exploiting the parametrization suggested there, the
function $Y^{(l)}(Q)$ turns out again to be dependent on three variables
$1\le\lambda_{1,2}\le\infty$ and $-1\le\lambda_0\le1$, and the analogue of
(\ref{l4}) reads
\begin{equation}\label{l5}\fl
\mathcal{F}_{\beta=1}(z',z'')=\int_1^{\infty}d\lambda_1
\int_1^{\infty}\rmd\lambda_2\int_{-1}^{1}\frac{\rmd\lambda_0}{\mathcal{R}^2}
\,\frac{\tilde{x}+\lambda_0 }{
\sqrt{(\tilde{x}+\tilde{\mu}_1)(\tilde{x}+\tilde{\mu}_2)}}\,\,
\,Y^{(l)}(\lambda_1,\lambda_2,\lambda_0)
\end{equation}
where
$\mathcal{R}=\lambda_0^2+\lambda_1^2+\lambda_2^2-2\lambda_0\lambda_1\lambda_2-1$
and $\tilde{\mu}_{1,2}=\lambda_1\lambda_2\pm
\sqrt{(\lambda_1^2-1)(\lambda_2^2-1)}$, and in zero dimension:
\begin{equation}\label{l7}
\fl Y^{(l)}(\lambda_1,\lambda_2,\lambda_0) = f(\{\lambda\})
e^{-\gamma(\lambda_1\lambda_2-\lambda_0)/2} \prod_{c=1}^{M}
\frac{g_c+\lambda_0}{\sqrt{(g_c+\tilde\mu_1)(g_c+\tilde\mu_2)}}\,.
\end{equation}
Here, $f(\{\lambda\})$ coming solely due to the $e^{-S_y\{Q\}}$ term is
explicitly given by (\ref{f}). Integration variables $\tilde\mu_{1,2}$ make
a connection between the ``radial'' parametrisation of $Q$ of
\cite{Efetov1983} and the ``angular'' parametrisation of
\cite{Verbaarschot1985}. This will be utilised in the next Appendix to get
explicit result for the zero-dimensional case for $\beta=1$.

\section{Analytic continuation (\ref{continuation}) and GOE result (\ref{Wgoe})}
We prove here equivalence of the analytic continuation of the ``connected''
part of $C(z',z'')$ in $z''\to-v\pm\rmi0$ (\ref{Phat}) to that of
$\mathcal{F}(\tilde{x})$ in $\tilde{x}\to-x\pm\rmi0$ (\ref{continuation}).
Let us consider in detail first the simplest case of $\beta=2$ symmetry
class in zero dimension. Then
\begin{eqnarray}
\fl C(z',z'') = \frac{1}{z'^2+(z''+1)^2}  - \frac{1}{4}\left(
\frac{\partial^2}{\partial z'^2} \!+\! \frac{\partial^2}{\partial z''^2}
\right) \nonumber
\\
\times\int_1^{\infty}\rmd\lambda_1\int_{-1}^{1}\frac{\rmd\lambda_0}
{\lambda_1-\lambda_0} \frac{2z''e^{-\gamma(\lambda_1-\lambda_0)/2} }{
z'^2+z''^2+2z''\lambda_1+1}
\end{eqnarray}
where we have kept in the integrand only the term remaining nonzero under
the differentiation. The FT $\hat{C}(k,z'')$ with respect to the first
argument reads as follows:
\begin{eqnarray}
\fl \widehat{C}(k,z'') = \frac{\pi e^{-|k|(z''+1)}}{z''+1} -
\frac{1}{4}\left( -k^2 \!+\! \frac{\partial^2}{\partial z''^2} \right)
\nonumber
\\
\times\int_1^{\infty}\rmd\lambda_1\int_{-1}^{1}\frac{\rmd\lambda_0}
{\lambda_1-\lambda_0} \frac{\pi e^{-|k|\sqrt{z''^2+2z''\lambda_1+1}} }{
\sqrt{z''^2+2z''\lambda_1+1}} 2z''e^{-\gamma(\lambda_1-\lambda_0)/2} \,.
\end{eqnarray}
Now we continue this result analytically in $z''$ and calculate the jump
(\ref{Phat}) on the negative real axis $z''\to-v\pm\rmi0$, with $v>0$. One
gets readily
\begin{eqnarray}
\fl \widehat{\mathcal{P}}(k,v) = \delta(v-1) + \frac{1}{2\pi}\left( -k^2
\!+\! \frac{\partial^2}{\partial v^2} \right) \nonumber
\\
\times\int_{\case{1+v^2}{2v}}^{\infty}\rmd\lambda_1\int_{-1}^{1}\frac{\rmd\lambda_0}
{\lambda_1-\lambda_0} \frac{ \cos(k\sqrt{2v\lambda_1-1-v^2}) }{
\sqrt{2v\lambda_1-1-v^2}} v\,e^{-\gamma(\lambda_1-\lambda_0)/2} \,.
\end{eqnarray}
Performing now the inverse FT one arrives at
\begin{eqnarray}\label{C1}
\fl \mathcal{P}(u,v) = \delta(u)\delta(v-1) +
\frac{1}{4\pi}\left(\frac{\partial^2}{\partial u^2} \!+\!
\frac{\partial^2}{\partial v^2} \right) \nonumber
\\
\times\int_{\case{1+v^2}{2v}}^{\infty}\rmd\lambda_1\int_{-1}^{1}\frac{\rmd\lambda_0}
{\lambda_1-\lambda_0} e^{-\gamma(\lambda_1-\lambda_0)/2}
\delta(u^2+v^2+1-2v\lambda_1)\,2v \,.
\end{eqnarray}
Making now use of
$\delta(u^2+v^2+1-2v\lambda_1)\,2v=\case{\partial}{\partial\lambda_1}
\Theta(2\lambda_1v-u^2-v^2-1)=\delta(\lambda_1-x)$, with
$x=\case1{2v}(u^2+v^2+1)$ from (\ref{P(u,v)}), one can immediately
integrate (\ref{C1}) over $\lambda_1$, yielding
$\int_{x-1}^{x+1}\rmd t\,t^{-1}e^{-\gamma t/2}$  for the second
line, that is in exact agreement with expression (\ref{Pgue})
following from the analytic continuation (\ref{continuation}).

The proof for the $\beta=1$ case proceeds along the same lines, explicit
expressions being however more lengthy. We omit it, considering instead the
less trivial derivation of expression (\ref{Wgoe}). It is important to
stress that for $\beta=1$ the nonzero $F(x)$ (\ref{Fcross}) at given $x>1$
is determined by the integration region
$\mathcal{B}_x=\{(\tilde{\mu}_{1},\tilde{\mu}_{2})\,
|\,x\leq\tilde{\mu}_1<\infty,\,0\leq\tilde{\mu}_2\leq x\}$, rather than by a
single point $\lambda_1=x$ as in the $\beta=2$ case. It is convenient,
therefore, to choose $\tilde{\mu}_{1,2}$ from (\ref{l5}) as new integration
variables. Actually, they are related to those from \cite{Verbaarschot1985}
(which are already used in (\ref{mu})) as follows:
\begin{equation}
\fl\mu_{1,2}=\case12(\tilde{\mu}_{1,2}-1)\equiv\case12\left[\lambda_1\lambda_2\pm
\sqrt{(\lambda_1^2-1)(\lambda_2^2-1)}\right]\,,\qquad
\mu_0=\case12(1-\lambda_0)\,.
\end{equation}
with the pre-exponential factor in (\ref{mu}) being the corresponding
integration measure. Expression (\ref{Fcross}) at $y=0$ acquires then the
following form (with $w\equiv\frac{x-1}{2}$):
\begin{eqnarray}\label{Fgoe}
\fl F(x)= \int_{0}^{1}\!\rmd\mu_0 \int_{w}^{\infty}\!\rmd\mu_1
\!\int_{0}^{w}\!\frac{\rmd\mu_2\,\,(1-\mu_0)\mu_0 (\mu_1-\mu_2) }{
\sqrt{\mu_1(1+\mu_1)\mu_2(1+\mu_2)} (\mu_0+\mu_1)^2
(\mu_0+\mu_2)^2 } \nonumber\\
\times e^{-\gamma(\mu_1+\mu_2+2\mu_0)/2}\frac{ w+\mu_0 }{
\sqrt{(\mu_1-w)(w-\mu_2)}}
\nonumber\\
\lo=\int_{0}^{1/w}\!\!\rmd\nu_0 \int_{1}^{\infty}\!\rmd\nu_1
\!\int_{0}^{1}\!\frac{\rmd\nu_2\,\,(1+\nu_0)\nu_0 (\nu_1-\nu_2) }{
\sqrt{\nu_1(\nu_1-1)\nu_2(\nu_2-1)} (\nu_0+\nu_1)^2
(\nu_0+\nu_2)^2 } \nonumber\\
\times e^{-\gamma w(\nu_1+\nu_2+2\nu_0)/2}\frac{ 1-w\nu_0 }{
\sqrt{(1+w\nu_1)(1+w\nu_2)}}
\end{eqnarray}
where the second equality comes after the scaling
$\mu_{0,1,2}=w\,\nu_{0,1,2}$.  It is convenient to consider now the function
$\frac{\rmd}{\rmd x}F(x)$. An essential observation is that this function
contains no contribution coming from the derivative of (\ref{Fgoe}) on the
upper limit due to vanishing of the integrand at $\nu_0=0,\case1w$. We
represent it in the following form:
\begin{equation}\label{Fgoe'}
\fl \frac{\rmd F(x)}{\rmd x} = \int_{0}^{1/w}\!\!\rmd\nu_0
\int_{1}^{\infty}\!\rmd\nu_1 \!\int_{0}^{1}\! \frac{\rmd\nu_2\,\, e^{-\gamma
w (\nu_1+\nu_2+2\nu_0)/2}\,X(\nu_0,\nu_1,\nu_2) }{
\sqrt{\nu_1(\nu_1-1)\nu_2(\nu_2-1)(1+w\nu_1)(1+w\nu_2)}}
\end{equation}
where a rational function $X$ contains all other terms resulting from the
differentiation. It is a crucial fact that $X$ can be further decomposed
into partial fractions with respect to $\nu_0$, yielding
\begin{equation}\label{Xgoe}
\fl X(\nu_0,\nu_1,\nu_2) = \sum_{i=1,2}(-1)^{i+1}\left[
\frac{a_i(\gamma+d_{12})}{(\nu_0+\nu_i)^2}+ \frac{\gamma b_i +c_id_{12} }{
\nu_0+\nu_i} + 2w\gamma\nu_i\right]
\end{equation}
with functions $a_i=\nu_i(\nu_i-1)(1+w\nu_i)$,
$b_i=1-2(1-w)\nu_i-3w\nu_i^2$, $c_i=1-2\nu_i-w\nu_i^2$ for $i=1,2$ and
$d_{12}=[(1+w\nu_1)(1+w\nu_2)]^{-1}$. Substituting (\ref{Xgoe}) in
(\ref{Fgoe'}), one readily sees that integrals over $\nu_0$ can be easily
performed while remaining integrations over $\nu_{1,2}$ get completely
decoupled in each term of the sum, leading finally to (\ref{Wgoe}).

\vspace*{2ex}
%

\end{document}